\documentclass[preprint,12pt]{elsarticle}
\biboptions{numbers,sort&compress}
\usepackage{amssymb}
\usepackage[margin=2.5cm]{geometry}
\usepackage{amsmath}
\usepackage{tabularx}
\usepackage[dvipsnames]{xcolor}
\usepackage{graphicx}
\usepackage{subcaption}
\usepackage{setspace}
\usepackage{placeins}
\usepackage[utf8]{inputenc}
\usepackage{pgfplots}
\usetikzlibrary{quotes,arrows.meta}
\usepgfplotslibrary{groupplots}
\usetikzlibrary{shapes.geometric}
\usepgfplotslibrary{groupplots,dateplot}
\usetikzlibrary{patterns,shapes.arrows}
\pgfplotsset{
  compat=newest
}

\usepackage{algorithm}
\usepackage{booktabs}
\usepackage{algpseudocode}
\usepackage{siunitx}
\usepackage{amsfonts}
\usepackage{bm}
\usepackage{caption}
\usepackage[hidelinks]{hyperref}
\usepackage{chngcntr}
\usepackage{booktabs,tabularx,array}
\usepackage{xcolor}

\begin{document}

\begin{frontmatter}

\title{A Sequential Monte Carlo Method for Data Assimilation in Subsurface Flow Systems with Hierarchical Uncertainty}
\author[a]{Wenchao Teng}
\author[a]{Guido Di Federico}
\author[a]{Louis J.~Durlofsky}

\affiliation[a]{{Department of Energy Science and Engineering, Stanford University, Stanford, CA 94305, USA}}
\begin{abstract}

Subsurface flow systems often involve hierarchical geological uncertainty, meaning both higher-level metaparameters and cell-by-cell properties are uncertain. This leads to challenges for data assimilation because both sets of parameters must be estimated. In this work we develop an adaptive-tempered sequential Monte Carlo (SMC) procedure for joint Bayesian estimation of these quantities. The method uses noncentered geological parameterizations, with geomodel realizations represented by metaparameters and Gaussian latent variables that are independent under the prior. Adaptive likelihood tempering, importance weighting, resampling, and blockwise Markov chain Monte Carlo (MCMC) mutation are applied at each iteration. Flow surrogates are used for all function evaluations. The hierarchical SMC procedure is further extended to treat successive data assimilation. In this case, existing posterior particles are updated in a manner that accounts for correlations in surrogate model errors across observation batches. We consider two distinct 3D systems with hierarchical uncertainty. The first is a faulted carbon storage system containing three stacked aquifers and two extensive faults. Principal component analysis is used to represent target-aquifer heterogeneity, while geological metaparameters characterize uncertain aquifer and fault properties. The second is a channelized oil-water system containing three facies with uncertain properties. A latent diffusion model represents variability in channel geometry and facies realizations. For the four synthetic true models considered, SMC posterior metaparameter distributions and flow predictions agree closely with reference MCMC results. SMC, however, requires fewer function evaluations and provides speedups of 2.5--7.0 for single-stage data assimilation. Even greater speedups are achieved with multiple GPUs.

\end{abstract}


\end{frontmatter}

\section{Introduction}
\label{sec:introduction}

The management of subsurface flow operations, such as geological carbon storage, oil production, and aquifer remediation, requires detailed numerical models that can reliably predict flow performance. To be predictive, these models must first be calibrated using observed data. This calibration problem is referred to as data assimilation or history matching. The large spatial scales over which subsurface flow processes operate, combined with the typically sparse data that may be collected at different times, pose challenges for data assimilation. These problems can be very high-dimension because the uncertain parameters, such as permeability and porosity, are represented block-by-block in large discretized models. In addition, in realistic systems geological uncertainty is often hierarchical, meaning the spatial property fields are themselves controlled by uncertain geological metaparameters (also referred to as hyperparameters or scenario parameters). Example metaparameters include the mean and standard deviation of log-permeability. In an overall sense, an effective data assimilation framework should be able to jointly estimate geological metaparameters and geomodel realizations in high-dimensional parameter spaces, update the models as new observations become available, and provide reliable posterior estimates and uncertainty quantification.

Consistent with these requirements, our focus in this work is on developing a sequential Monte Carlo (SMC) procedure for data assimilation in subsurface flow systems with hierarchical uncertainty. The procedure jointly estimates geological metaparameters and latent variables defining geomodel realizations, and enables efficient successive data assimilation as new observations become available. We consider two distinct subsurface flow settings: a faulted carbon storage system with three stacked aquifers and two large faults, and a channelized oil-water system with injection and production wells. Both problems are 3D, high-dimensional, and characterized by hierarchical uncertainty. Geological parameterization techniques -- namely principal component analysis (PCA) for the faulted system and latent diffusion models (LDMs) for the channelized system -- are used to reduce dimensionality and enable dimension-robust sampling of latent Gaussian variables while retaining geological realism. Because the data assimilation procedure requires a large number of forward evaluations, deep learning surrogate models are applied in place of high-fidelity flow simulations. The performance of our SMC procedure is assessed through extensive comparisons with reference Markov chain Monte Carlo (MCMC) results.

Data assimilation has been widely applied to subsurface flow problems, and our review of previous work will focus on the most popular and most relevant methods. Ensemble Kalman methods are attractive because they are derivative-free and computationally efficient~\citep{evensen2003Ensemble,aanonsen2009,evensen2000ensemble,emerick2013Ensemble,Huang2022}. In traditional applications, prior geomodel realizations are generated using a fixed set of geological metaparameters. In reality, however, these quantities may be highly uncertain and can strongly influence the spatial structure and variation of prior models. Hierarchical Bayesian formulations address this issue by assigning hyperprior distributions to the metaparameters and estimating them together with realization variables. Several ensemble-based extensions have been developed for such problems. \citet{emerick2016Hierarchical} introduced a treatment of uncertainty in the prior mean and covariance inspired by Gaussian mixtures. \citet{chada2018Parameterizations} investigated hierarchical parameterizations for uncertain spatial structure. \citet{oliver2022Hybrid} developed a hybrid iterative ensemble smoother for uncertain covariance parameters. These approaches retain the computational advantages and natural parallelization of ensemble methods. Their Kalman-based updates, however, rely on finite ensemble covariance estimates and statistical linearization, so the resulting ensembles are not in general exact samples from the target Bayesian posterior for nonlinear and non-Gaussian problems~\citep{ernst2015Analysis,emerick2013Sampling,ClearyEtAl2021}. This issue can be particularly important for systems with hierarchical uncertainty~\citep{oliver2022Hybrid,teng2025likelihood}. Limitations associated with ensemble Kalman methods when used in combination with LDM-based parameterizations have also been identified~\citep{difederico2026data}. In that study, these methods were shown to significantly overestimate  posterior uncertainty relative to formal sampling procedures for channelized systems parameterized using LDMs.

An alternative data assimilation approach is to apply formal Monte Carlo sampling techniques that target the Bayesian posterior. Along these lines, MCMC methods designed for high-dimensional parameter spaces have been used for hierarchical inference problems in subsurface systems~\citep{hansen2012Inverse,reuschen2020Bayesian,xiao2021Bayesian,han2024Surrogate}. These approaches can generate asymptotically exact posterior samples for nonlinear and non-Gaussian problems without the approximations used in ensemble Kalman methods. They may, however, require very long Markov chains to converge and generate a sufficient number of effective posterior samples. This can lead to computational challenges since these methods are largely serial, although modest degrees of parallelization can be achieved using multiple chains. SMC methods provide a population-based alternative~\citep{liu1998sequential,DelMoral2006,chopin2020introduction} with some conceptual advantages over MCMC procedures. SMC samplers, which parallelize naturally, transport a collection of weighted particles through a sequence of intermediate distributions bridging the prior and posterior. These methods have been developed for high-dimensional inverse problems~\citep{Koutsourelakis2009,kantas2014sequential,beskos2015Sequential,latz2018Multilevel}. SMC has also been applied to geophysical inversion~\citep{amaya2021adaptive} and groundwater problems~\citep{ellam2016bayesian,amaya2022hydrogeological}. To our knowledge, however, previous SMC studies have not addressed the joint estimation of geological metaparameters and latent variables in subsurface flow settings, nor have they considered successive data assimilation for these problems.

In recent work, we presented an SMC-based approximate Bayesian computation (SMC-ABC) method for geological metaparameter estimation~\citep{teng2025likelihood}. SMC-ABC is a widely used likelihood-free sampling approach for Bayesian inference problems characterized by intractable likelihood functions~\citep{sisson2007sequential,beaumont2009adaptive,toni2009approximate}. Likelihood-free methods have been applied in a range of scientific and engineering fields, though their use in subsurface inverse problems remains limited~\citep{vrugt2013toward,cui2018emulator,teng2025likelihood}. Despite their similar-sounding names, the SMC-ABC approach applied in~\citep{teng2025likelihood} and the method presented here differ fundamentally in their estimation target and treatment of hierarchical uncertainty. The SMC-ABC procedure decomposes the hierarchical problem into the marginal inference of geological metaparameters, followed by conditional geomodel estimation using an ensemble Kalman method. The likelihood-free formulation was used because computation of the marginal likelihood would require integration over all possible geological realizations, which is computationally intractable. In the ``centered'' representation used in the SMC-ABC formulation, geomodel realizations are conditioned on the geological metaparameters, and the two can be strongly dependent a priori. The geomodel realizations are not expressed in terms of latent variables; rather, they must be constructed using geostatistical software for each proposed set of metaparameters. Thus, our previous SMC-ABC methodology is appropriate for cases in which a geological parameterization of the cell-by-cell properties is not available.

The SMC procedure proposed in this work uses ``noncentered'' geomodel representations, in which each realization is specified by geological metaparameters and latent (e.g., PCA or LDM) variables. These two sets of variables are independent under the prior, though they may become dependent in the posterior~\citep{Papaspiliopoulos2007}. Our SMC procedure jointly estimates the geological metaparameters and latent variables within a single likelihood-based framework. The method derives advantages from the Gaussian character of the latent variables. This is a useful feature since many parameterization procedures, including those applied here as well as other popular methods like variational autoencoders (VAEs), construct a latent space that is approximately multi-Gaussian. The proposed methodology also avoids the approximations introduced by the ABC discrepancy threshold and the Kalman update required in the previous SMC-ABC approach.

The SMC procedure developed here provides efficient hierarchical Bayesian inference by transporting a population of particles from the joint prior to the hierarchical posterior through importance weighting, resampling, and short-chain MCMC mutation. We believe this to be the first such method developed and applied for subsurface flow applications. In our procedure, adaptive tempering is used to incorporate the likelihood progressively and thus limit particle weight degeneracy. At intermediate tempering iterations, resampling is followed by short blockwise MCMC chains, with preconditioned Crank--Nicolson (pCN) proposals for the Gaussian latent variables and Gaussian random-walk proposals for the geological metaparameters. The resulting weighted particles provide a joint posterior approximation for geological metaparameters and geomodel latent variables. The performance of this procedure is evaluated through detailed comparisons with reference MCMC results for the faulted carbon storage and channelized oil-water systems. 
Efficiency metrics that demonstrate the advantages of SMC over MCMC are provided. The population-based structure of SMC enables additional parallelization over multiple GPUs, and the further benefits achieved with this treatment are quantified.

In real applications, data often become available at a series of times as operations progress. Along these lines, another new contribution in this work is the extension of the hierarchical SMC procedure to treat successive (multistage) data assimilation. This is accomplished by updating the existing joint posterior particles through a new tempering sequence based on the conditional likelihood of the additional observations. This conditional likelihood accounts for correlations in surrogate model errors across observation batches, while the likelihood of previously assimilated observations remains fully incorporated in the tempered target distribution. We assess the successive SMC procedure for the faulted carbon storage system through comparisons with single-stage SMC and MCMC calculations. We reiterate that the overall SMC method presented in this study does not ``replace'' our previous SMC-ABC approach, since the two methodologies are applicable for different types of hierarchical descriptions.

This paper proceeds as follows. In Section~\ref{sec:sim_setup}, we describe the two subsurface flow systems considered in this study, along with the geological parameterizations and deep learning-based surrogate models used in the data assimilation procedure. A few of the subsurface flow surrogate models most relevant to this study are also discussed in Section~\ref{sec:sim_setup}. We present the hierarchical Bayesian formulation, the adaptive tempered SMC method, and the procedure for successive data assimilation in Section~\ref{sec:smc_hierarchical}. The MCMC method used to provide reference posterior results is described in Section~\ref{sec:mcmc}. Detailed data assimilation results for the faulted carbon storage system are presented in Section~\ref{sec:carbon_results}, and the corresponding results for the channelized oil-water system appear in Section~\ref{sec:oilwater_results}. We conclude in Section~\ref{sec:conclusions} with a summary of our main findings and suggestions for future work. Additional details and data assimilation results are provided in online Supplementary Material.

\section{Geomodels, simulation setup, and surrogate models}
\label{sec:sim_setup}

In this section, we describe the two subsurface flow systems used to assess the capabilities of SMC for hierarchical data assimilation. The first is a 3D carbon storage system with faults, and the second is a 3D oil-water system with channelized geomodels. For each case, we present the geomodel and geological parameterization, followed by the flow simulation setup and the deep learning surrogate model used for function evaluations. Many of the procedures described here for geomodeling and flow surrogate modeling are the same as in previous studies, though there are important differences in some treatments (which will be noted).

\subsection{Carbon storage system with faults}
\label{sec:co2_geomodel}

\begin{figure}[!htbp]
\centering
\includegraphics[width=\linewidth]
{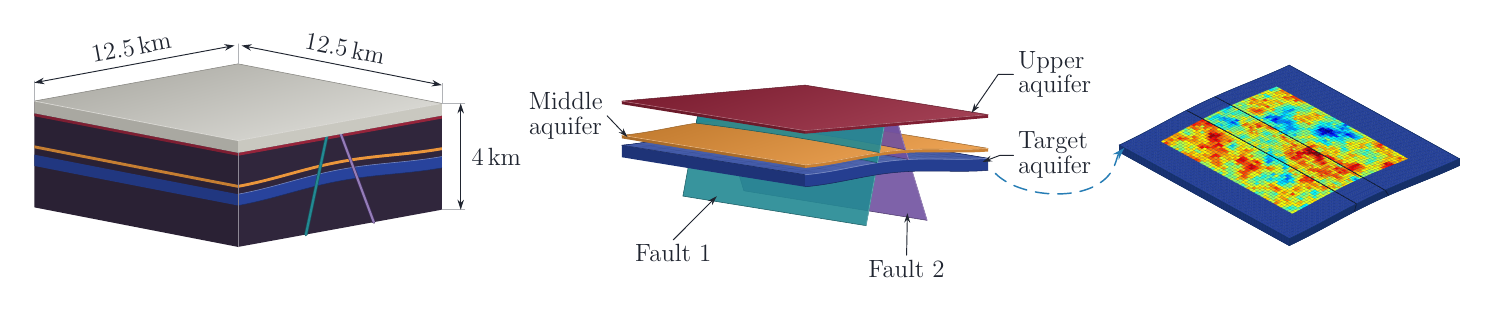}
\caption{Overall domain of the modified SEAM carbon storage geomodel (left image), the three stacked aquifers and two faults (middle image), and storage aquifer including the surrounding region (right image).}
\label{fig:seam_domain}
\end{figure}

The faulted carbon storage system considered in this study is based on the modified SEAM CO$_2$ geomodel developed by~\citet{HanDurlofsky2026}. As shown in Fig.~\ref{fig:seam_domain}, the model includes a target aquifer, two overlying aquifers separated by low-permeability caprock, surrounding formations, and two extensive faults that intersect all three aquifers. The target aquifer, where CO$_2$ is injected, is $8.1$~km $\times$ $9.7$~km $\times$ 425~m and is discretized on a $48\times48\times25$ grid. The middle aquifer spans $12.5$~km $\times$ $12.5$~km $\times$ 102.5~m and is represented on a $69\times62\times6$ grid. Similar dimensions and grid resolution are used for the upper aquifer. The full numerical model contains $69\times62\times78=333{,}684$ grid cells. The two faults are separated by about 2~km within the target aquifer and extend through the overlying formations. Depending on their permeabilities, they can provide pathways for pressure communication and upward CO$_2$ migration between the aquifers.

Geological uncertainty in the faulted carbon storage system is represented hierarchically through 11 metaparameters, denoted $\mathbf{h}\in\mathbb{R}^{11}$, and realization-specific spatial variability within the target aquifer. The five target-aquifer metaparameters are the mean and standard deviation of log-permeability, $\mu_{\log k}$ and $\sigma_{\log k}$, the permeability anisotropy ratio $a_r$, and the coefficients $d$ and $e$ relating porosity and log-permeability (the exact expression is given below). The remaining metaparameters are the permeabilities of the middle and upper aquifers, $k_m$ and $k_u$, and four fault-segment permeabilities. For each fault $j=1,2$, $k_{fj}^{tm}$ denotes the permeability of the segment connecting the target and middle aquifers (thus the `tm' superscript), and $k_{fj}^{mu}$ indicates the permeability of the segment connecting the middle and upper aquifers (`mu' superscript). This treatment allows fault permeability to vary with depth, which is geologically realistic. The prior distributions of the 11 metaparameters are given in Table~\ref{tab:carbon_meta_priors}.

\begin{table}[!htbp]
\small
\centering
\caption{Prior distributions of the geological metaparameters for the faulted carbon storage system.}
\label{tab:carbon_meta_priors}
\begin{tabular}{p{9.0cm} p{5.0cm}}
\hline
\noalign{\hrule height 0.8pt}
Metaparameter & Prior distribution \\
\hline
\multicolumn{2}{l}{\textbf{Target aquifer}} \\
Mean of log-permeability, $\mu_{\log k}$ & $\mu_{\log k}\sim U(4,6)$ \\
Standard deviation of log-permeability, $\sigma_{\log k}$ & $\sigma_{\log k}\sim U(1,1.5)$ \\
Permeability anisotropy ratio, $a_r$ & $\log_{10}a_r\sim U(-1.3,-0.7)$ \\
Coefficient in log-permeability--porosity expression, $d$ & $d\sim U(0.02,0.04)$ \\
Coefficient in log-permeability--porosity expression, $e$ & $e\sim U(0.06,0.08)$ \\
\hline
\multicolumn{2}{l}{\textbf{Overlying aquifers}} \\
Middle-aquifer permeability, $k_m$ & $k_m\sim U(50,500)$~md \\
Upper-aquifer permeability, $k_u$ & $k_u\sim U(50,500)$~md \\
\hline
\multicolumn{2}{l}{\textbf{Faults}} \\
Fault 1 target--middle permeability, $k_{f1}^{tm}$ & $\log_{10}k_{f1}^{tm}\sim U(-1,2.5)$ \\
Fault 1 middle--upper permeability, $k_{f1}^{mu}$ & $\log_{10}k_{f1}^{mu}\sim U(-1,2.5)$ \\
Fault 2 target--middle permeability, $k_{f2}^{tm}$ & $\log_{10}k_{f2}^{tm}\sim U(-1,2.5)$ \\
Fault 2 middle--upper permeability, $k_{f2}^{mu}$ & $\log_{10}k_{f2}^{mu}\sim U(-1,2.5)$ \\
\hline
\noalign{\hrule height 0.8pt}
\end{tabular}
\end{table}

The spatial heterogeneity of the target aquifer is represented using principal component analysis (PCA). The PCA basis is constructed from 1000 standard multi-Gaussian realizations, with the prescribed spatial correlation structure, generated using SGeMS~\citep{Remy2009}. Spatial correlations are incorporated using a spherical variogram with correlation lengths of about 1700~m, 2000~m and 50~m (in the $x$, $y$ and $z$-directions). The first $n_\xi=820$ principal components are retained, preserving approximately 95\% of the total energy. A realization of the standard multi-Gaussian field, $\mathbf{y}_{\mathrm{pca}}\in\mathbb{R}^{n_c}$, is expressed as
\begin{equation}
\mathbf{y}_{\mathrm{pca}}=\boldsymbol{\Phi}\boldsymbol{\xi}+\overline{\mathbf{y}},
\end{equation}
where $n_c=57{,}600$ is the number of target-aquifer cells, $\boldsymbol{\Phi}\in\mathbb{R}^{n_c\times n_\xi}$ is the truncated PCA basis matrix, $\boldsymbol{\xi}\in\mathbb{R}^{n_\xi}$ is the latent variable vector, and $\overline{\mathbf{y}}\in\mathbb{R}^{n_c}$ is the mean of the realizations used to construct the basis. The horizontal permeability $k_{h,i}$, vertical permeability $k_{v,i}$, and porosity $\phi_i$ in cell $i$ are computed from the metaparameters and PCA variables via
\begin{equation}
k_{h,i}=\exp\left(\sigma_{\log k}y_{\mathrm{pca},i}+\mu_{\log k}\right),\quad
k_{v,i}=a_r k_{h,i},\quad
\phi_i=d\log(k_{h,i})+e,\quad i=1,\ldots,n_c.
\end{equation}
We reiterate that the variables $\mathbf{h}$ and $\boldsymbol{\xi}$ fully specify the geological realization used in the forward model.

Supercritical CO$_2$ is injected into the storage aquifer through two vertical wells that are perforated (open to flow) over the full target aquifer thickness. Each well operates at a constant injection rate of 1~Mt/year for 50~years, resulting in a total injected mass of 100~Mt. These rates are sufficiently low such that bottom-hole pressure constraints are not triggered. High-fidelity flow simulations are performed using the GEOS simulator~\citep{Settgast2024}, with no-flow boundary conditions imposed along the boundaries of the model domain. Relative permeability hysteresis, capillary pressure, and effective rock compressibility are included. The initial pressure and temperature conditions vary with depth, with values of 18~MPa and $67\,^{\circ}$C specified at a depth of 1880~m in the target aquifer. For additional details on the simulation specifications, please refer to~\citet{HanDurlofsky2026}. The GEOS simulations require an average wall-clock time of about 7~minutes using 32 CPU cores on AMD EPYC compute nodes.

We apply deep learning surrogate models to reduce the cost of the large number of flow evaluations required during data assimilation. For geological carbon storage, these models commonly provide parameter-to-state mappings, in our case from the cell-by-cell geological property fields to time-dependent saturation and pressure fields. Architectures considered for this purpose include recurrent residual U-Nets~\citep{tang2022Deep,han2024Surrogate}, Fourier neural operators~\citep{wen2022UFNO,witte2023Fast,tang2024Multifidelity}, transformer-based networks~\citep{seabra2024AI,HanDurlofsky2026}, and graph neural networks~\citep{tang2025Graph}. For the faulted system considered here, we use (with slight modification) the recurrent transformer U-Net model introduced by \citet{HanDurlofsky2026}. This architecture combines a residual U-Net encoder--decoder, a transformer block, and a 3D convolutional long short-term memory network to capture long-range spatial effects along with temporal evolution. In this study the transformer block is placed before the residual block at the $12\times12\times12$ encoder level, rather than at the end of the encoder as in \citet{HanDurlofsky2026}. This alternate configuration was suggested (though not applied) in the earlier study, and we found it to provide slightly more accurate results here. We use separate networks to predict CO$_2$ saturation and pressure at 11 prespecified time steps, $t=2$, 4, 6, 8, 10, 12, 16, 20, 30, 40, and 50~years. These time steps are different than those in \citet{HanDurlofsky2026} and were selected for use in our successive data assimilation example.

A total of 5000 high-fidelity simulations are performed for realizations generated from randomly sampled metaparameters $\bf h$ and PCA variables $\boldsymbol{\xi}$. Of these, 4000 simulations are used for training, 500 for validation, and 500 for testing. The training procedure and error metrics follow~\citet{HanDurlofsky2026}. Each network is trained for 300 epochs with batch sizes of 4 for saturation and 8 for pressure. Training requires about 23 and 21~hours for the saturation and pressure networks, respectively, on a single NVIDIA A100 GPU with 80~GB of memory. Over the 500 test-set realizations, the median saturation mean absolute error is 0.028 and the median relative pressure error is 0.08\%. These values are comparable to the corresponding median errors of 0.025 and 0.12\% reported in~\citep{HanDurlofsky2026} for a similar (though not identical) setup.

\begin{figure}[!htbp]
\centering
\includegraphics[width=0.95\linewidth]
{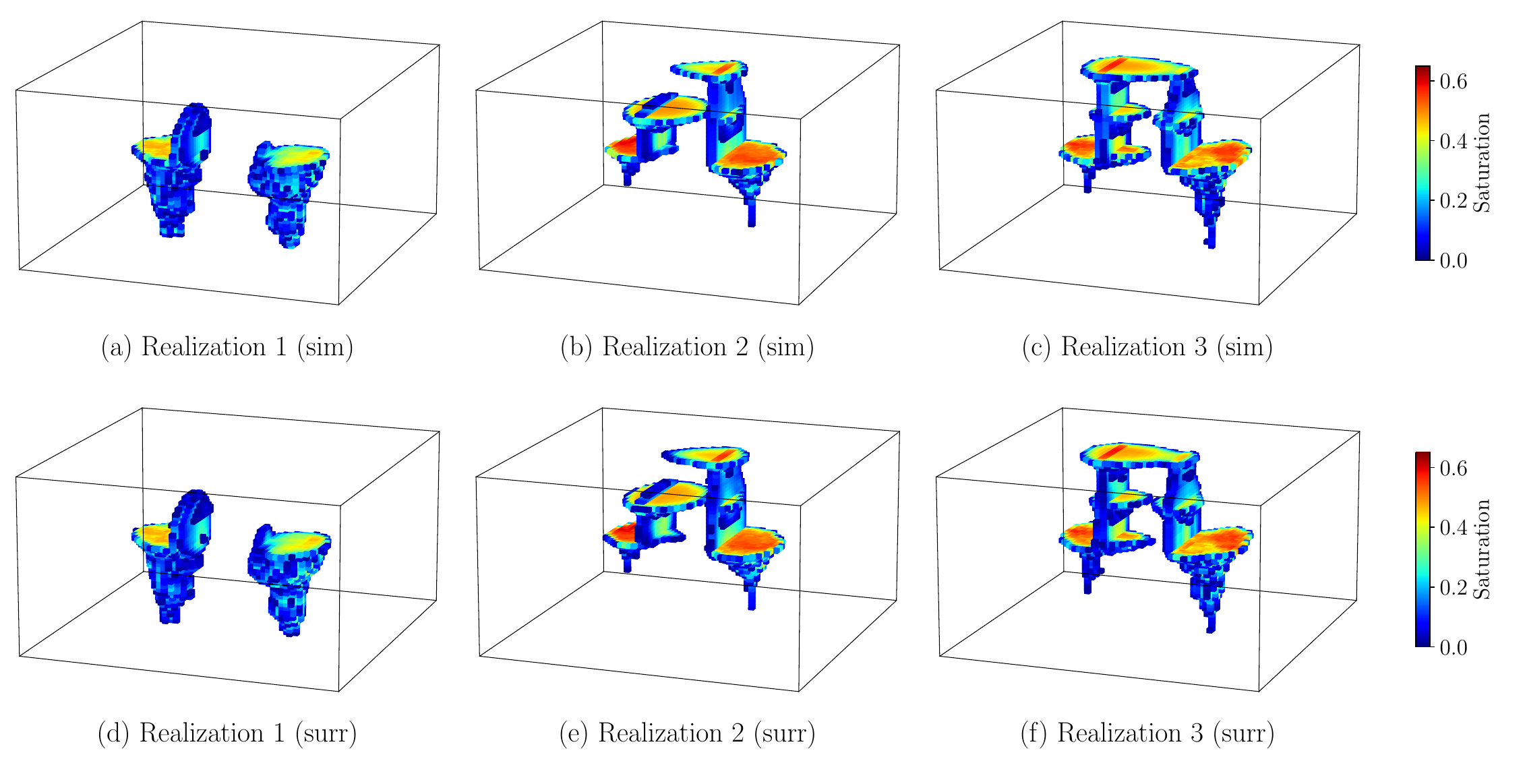}
\caption{$\mathrm{CO}_2$ saturation over the full domain at 50~years for three test-set realizations. GEOS flow simulation results are shown in the top row and surrogate model predictions in the bottom row.}
\label{fig:saturation_comparison}
\end{figure}

Fig.~\ref{fig:saturation_comparison} shows 3D CO$_2$ saturation fields at 50~years from flow simulations (top row) and from corresponding surrogate predictions (bottom row) for three test-set realizations with distinct fault-leakage behaviors. In realization~1, the target-to-middle permeabilities of both faults are low ($k_{f1}^{tm}=1.9$~md and $k_{f2}^{tm}=0.34$~md). Thus, CO$_2$ remains largely within the target aquifer, with only limited migration up the faults. In realization~2, the target-to-middle fault segments are highly permeable ($k_{f1}^{tm}=174$~md and $k_{f2}^{tm}=126$~md), as is the middle-to-upper permeability of fault~2 ($k_{f2}^{mu}=148$~md). The corresponding permeability for fault~1 (leftmost fault) is much lower ($k_{f1}^{mu}=3.7$~md). As a result, CO$_2$ reaches both overlying aquifers through fault~2, while migration along fault~1 is largely confined to the middle aquifer. In realization~3, all four fault segments have relatively high permeabilities, ranging from 56 to 182~md, and CO$_2$ migrates to the upper aquifer through both faults. In all three cases, despite the qualitatively different flow behaviors, the surrogate closely reproduces the plume geometries and lateral spreading observed in the high-fidelity simulations. Comparisons of pressure from flow simulations and surrogate predictions are provided in Fig.~S1 in Supplementary Material. High accuracy in the surrogate results is again observed.

\FloatBarrier

\subsection{Oil-water system with channelized geomodels}
\label{sec:oil_geomodel}

The channelized geomodels considered in the oil-water case contain three facies (rock types), sand/channel, mud/shale, and levee. The facies system is the same as that in \citet{difederico2026data}, but here we also introduce uncertain facies-dependent petrophysical properties (permeability and porosity for each rock type). This system actually contains three levels of uncertainty, because we have uncertain channel geometry parameters, uncertain channel realizations for the given geometric parameters, and uncertain facies-dependent properties for each realization. However, because the 3D-LDM parameterization used here represents both the channel geometry and channel realizations with a single latent variable, the data assimilation algorithms treat only two levels of hierarchical uncertainty, as in the carbon storage example.

\begin{figure}[!htbp]
\centering
\includegraphics[width=0.95\linewidth]
{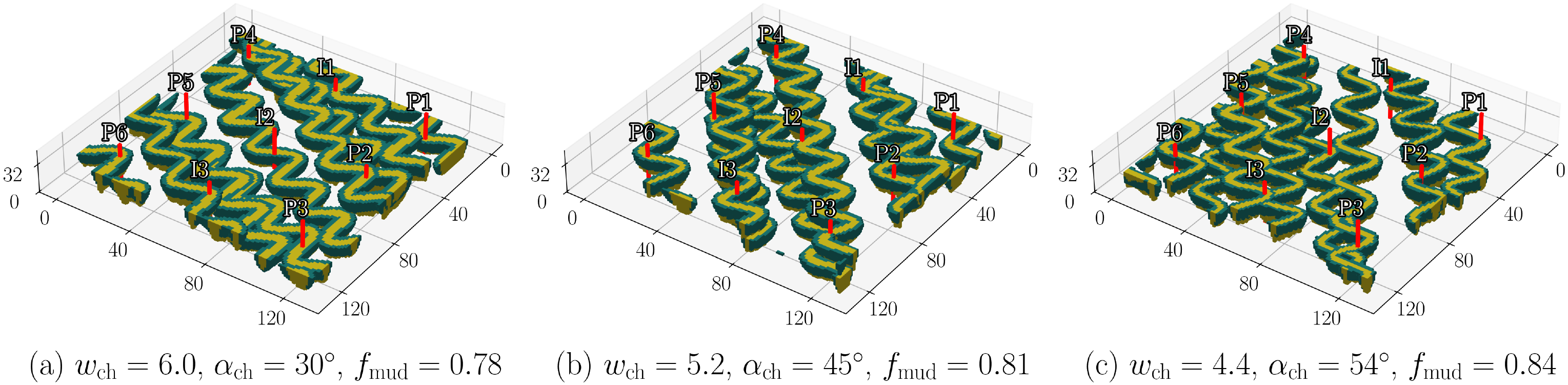}
\caption{Representative facies realizations generated using Petrel, with background mud omitted. Corresponding geometric metaparameters (channel width, channel orientation, and mud fraction) are given. Injectors (I1--I3) and producers (P1--P6) are also shown.}
\label{fig:examples_petrel}
\end{figure}

The geometric metaparameters are denoted $\mathbf{h}_g=[w_{\mathrm{ch}},\alpha_{\mathrm{ch}},f_{\mathrm{mud}}]$, where $w_{\mathrm{ch}}$ is the average channel width (in units of grid blocks), $\alpha_{\mathrm{ch}}$ is average channel orientation, and $f_{\mathrm{mud}}$ is mud fraction. To construct a random channelized geomodel, we draw samples of $w_{\mathrm{ch}}$, $\alpha_{\mathrm{ch}}$ and $f_{\mathrm{mud}}$ from the prior ranges given in Table~\ref{metaparameter_priors} and then apply object-based modeling in Petrel~\citep{Schlumberger} to generate a realization for the sampled $\mathbf{h}_g$. A total of 3000 such facies realizations are constructed to train the 3D-LDM. Each realization, denoted by $\mathbf{m}\in\mathbb{R}^{n_c}$, is defined on a $128\times128\times32$ grid (for a total of $n_c=524{,}288$ cells). All realizations are conditioned to the same hard facies data at three injector (I1--I3) and six producer (P1--P6) locations. Representative realizations corresponding to different values of $\mathbf{h}_g$, along with the nine wells, are shown in Fig.~\ref{fig:examples_petrel}. In this figure, we omit the background mud facies to better display the 3D channel and levee structures.

The channelized geomodels are parameterized using the 3D latent diffusion model (3D-LDM) developed by~\citet{difederico2025ThreeDimensional}. The 3D-LDM represents the high-dimensional facies field using a low-dimensional latent variable, $\boldsymbol{\xi}_T\in\mathbb{R}^{n_\xi}$, with $n_\xi=1024$. A realization of $\boldsymbol{\xi}_T$ is mapped through a sequence of denoising steps and the decoder of a variational autoencoder to generate the corresponding facies realization, $\mathbf{m}_0^{\mathrm{LDM}}(\boldsymbol{\xi}_T)$. In \citet{difederico2025ThreeDimensional}, ensembles of Petrel and 3D-LDM realizations were shown to be in close agreement in terms of experimental variograms, geometric metaparameter distributions, and flow responses. These results demonstrate that the 3D-LDM provides a lower-dimensional representation while preserving the essential features of the channelized geomodels. We reiterate that a single 3D-LDM parameterization is applicable over the full prior ranges (given in Table~\ref{metaparameter_priors}) for $w_{\mathrm{ch}}$, $\alpha_{\mathrm{ch}}$ and $f_{\mathrm{mud}}$.

In contrast to previous studies~\citep{difederico2025ThreeDimensional,difederico2026data}, the petrophysical metaparameters are now treated as uncertain. The uncertain parameters are \[ \mathbf{h}_p=[(\phi_{\mathrm{ch}},k_{\mathrm{ch}},a_{r,\mathrm{ch}}),(\phi_{\mathrm{lv}},k_{\mathrm{lv}},a_{r,\mathrm{lv}}),(\phi_{\mathrm{mud}},k_{\mathrm{mud}},a_{r,\mathrm{mud}})], \] where $\phi$, $k$, and $a_r$ denote porosity, horizontal permeability, and permeability anisotropy ratio, respectively, and the subscripts $\mathrm{ch}$, $\mathrm{lv}$, and $\mathrm{mud}$ refer to the channel, levee, and mud facies. The prior ranges for the variables in $\mathbf{h}_p$ are given in Table~\ref{metaparameter_priors}. Although $\mathbf{h}_p$ varies from realization to realization, for a particular realization the set of properties defined by $\mathbf{h}_p$ are taken to be uniform within each facies. The introduction of uncertain $\mathbf{h}_p$ represents an important difference from our earlier studies, where a channel-to-mud permeability ratio of 40 and a uniform permeability anisotropy ratio of 0.2 were applied~\citep{difederico2025ThreeDimensional,difederico2026data}. The priors in this study span permeability ratios from 50 to 500, and anisotropy ratios as low as 0.01 in the levee and mud facies. These broader ranges and additional uncertainties lead to greater variability in the flow response and introduce significant challenges for the flow surrogate.

\begin{table}[!htbp]
\small
\centering
\caption{Prior distributions of the metaparameters for the channelized oil-water system.}
\label{metaparameter_priors}
\begin{tabular}{p{9.5cm} p{5.0cm}}
\hline
\noalign{\hrule height 0.8pt}
Metaparameter & Prior distribution \\
\hline
\multicolumn{2}{l}{\textbf{Geometric metaparameters}} \\
Average channel width, $w_{\mathrm{ch}}$ & $w_{\mathrm{ch}}\sim U(4,6)$ cells \\
Average channel orientation, $\alpha_{\mathrm{ch}}$ & $\alpha_{\mathrm{ch}}\sim U(30^\circ,60^\circ)$ \\
Mud fraction, $f_{\mathrm{mud}}$ & $f_{\mathrm{mud}}\sim U(0.72,0.87)$ \\
\hline
\multicolumn{2}{l}{\textbf{Petrophysical metaparameters}} \\
Channel porosity, $\phi_{\mathrm{ch}}$ & $\phi_{\mathrm{ch}}\sim U(0.21,0.30)$ \\
Channel horizontal permeability, $k_{\mathrm{ch}}$ & $k_{\mathrm{ch}}\sim U(500,1500)$~md \\
Channel permeability anisotropy ratio, $a_{r,\mathrm{ch}}$ & $a_{r,\mathrm{ch}}\sim U(0.10,0.50)$ \\
Levee porosity, $\phi_{\mathrm{lv}}$ & $\phi_{\mathrm{lv}}\sim U(0.11,0.20)$ \\
Levee horizontal permeability, $k_{\mathrm{lv}}$ & $k_{\mathrm{lv}}\sim U(50,300)$~md \\
Levee permeability anisotropy ratio, $a_{r,\mathrm{lv}}$ & $a_{r,\mathrm{lv}}\sim U(0.01,0.10)$ \\
Mud porosity, $\phi_{\mathrm{mud}}$ & $\phi_{\mathrm{mud}}\sim U(0.01,0.10)$ \\
Mud horizontal permeability, $k_{\mathrm{mud}}$ & $k_{\mathrm{mud}}\sim U(3,10)$~md \\
Mud permeability anisotropy ratio, $a_{r,\mathrm{mud}}$ & $a_{r,\mathrm{mud}}\sim U(0.01,0.10)$ \\
\hline
\noalign{\hrule height 0.8pt}
\end{tabular}
\end{table}

The facies realization and sampled petrophysical metaparameters together specify the geomodel used in reservoir simulation. Two-phase immiscible (oil-water) flow simulations are performed using the tNavigator simulator~\cite{rock_flow_dynamics}. The simulation period is 1500~days. All wells operate under bottom-hole pressure control and are perforated only in channel-facies cells. Well rates are reported at 50-day intervals, giving 30 reporting times. Each simulation run requires an average of about 8~minutes using 16 cores on an AMD EPYC compute node. Additional information on the channelized facies and simulation specifications can be found in~\citet{difederico2025ThreeDimensional}.

As in the faulted carbon storage model, although the time needed for a single run is modest, the large number of function evaluations required for SMC and MCMC necessitates a flow surrogate model. For this class of problems, deep learning surrogates are often constructed as direct parameter-to-observable mappings because the quantities of interest are well-level production responses (i.e., these models do not provide pressure and saturation in every cell). Image-to-sequence architectures typically use convolutional networks to encode high-dimensional property fields and recurrent networks to map the extracted spatial features to well-rate time series~\citep{ma2022Efficient,difederico2026data}. This architecture can also incorporate dynamic well control schedules as recurrent input for applications in closed-loop reservoir management~\citep{kim2023Convolutional}.

The surrogate developed in this study follows the general convolutional--recurrent approach of~\citet{difederico2026data}. Here, however, we extend the input representation and spatial feature processing to account for uncertain facies-dependent petrophysical properties. The earlier model encodes the facies field into a single globally pooled vector and then uses a long short-term memory network for temporal decoding. Here, a 3D residual convolutional encoder processes the facies geometry and associated petrophysical property fields, while a separate encoder represents the petrophysical metaparameters. Attention components then incorporate reservoir-scale information and features from regions connecting injector--producer pairs. These features are combined with the encoded metaparameters to construct well-specific representations. A subsequent transformer block enables information exchange among the wells. A gated recurrent unit captures temporal evolution to predict well rates at the 30 reporting times. The detailed architecture of the new flow surrogate is provided in Table~S1.

The training set for the flow surrogate comprises 9000 tNavigator simulations for geomodels that cover the range of $\mathbf{h}_g$ and $\mathbf{h}_p$. These geomodels are obtained by assigning three Latin-hypercube samples of $\mathbf{h}_p$, from the prior ranges given in Table~\ref{metaparameter_priors}, to each of the 3000 Petrel geomodels used for 3D-LDM training. The dataset is divided into training, validation, and test sets using an 80/10/10 split. All three petrophysical realizations associated with a given geomodel are retained within the same subset. Thus, the validation and test sets contain facies realizations not seen during training. The network is trained for 200~epochs, with the goal of minimizing discrepancies between predicted and simulated well-rate time series. An initial learning rate of 0.0003 is used, with a five-epoch warm-up followed by cosine decay. Training requires about 7.6~hours on a single NVIDIA A100 GPU with 40~GB of memory.

We evaluate the surrogate on the test set using the same relative error metrics as in \citet{difederico2026data}. The median relative errors are 4.8\% for oil production, 8.1\% for water production, and 4.6\% for water injection. These errors are comparable to, but slightly larger than those in the earlier study (where the median errors were 4.6\%, 6.1\%, and 3.4\%, respectively). The larger errors here are due to the broad range and increased variability in flow responses associated with the uncertain $\mathbf{h}_p$ metaparameters. Surrogate prediction errors are higher for large channel-to-mud permeability contrasts. Further improvement in surrogate accuracy could be achieved through active learning, with additional high-fidelity training simulations targeted to regions of the petrophysical parameter space associated with larger errors. This was not attempted here, however, because the flow surrogate is already sufficiently accurate for our purposes.

\begin{figure}[!htbp]
\centering
\includegraphics[width=0.95\linewidth]
{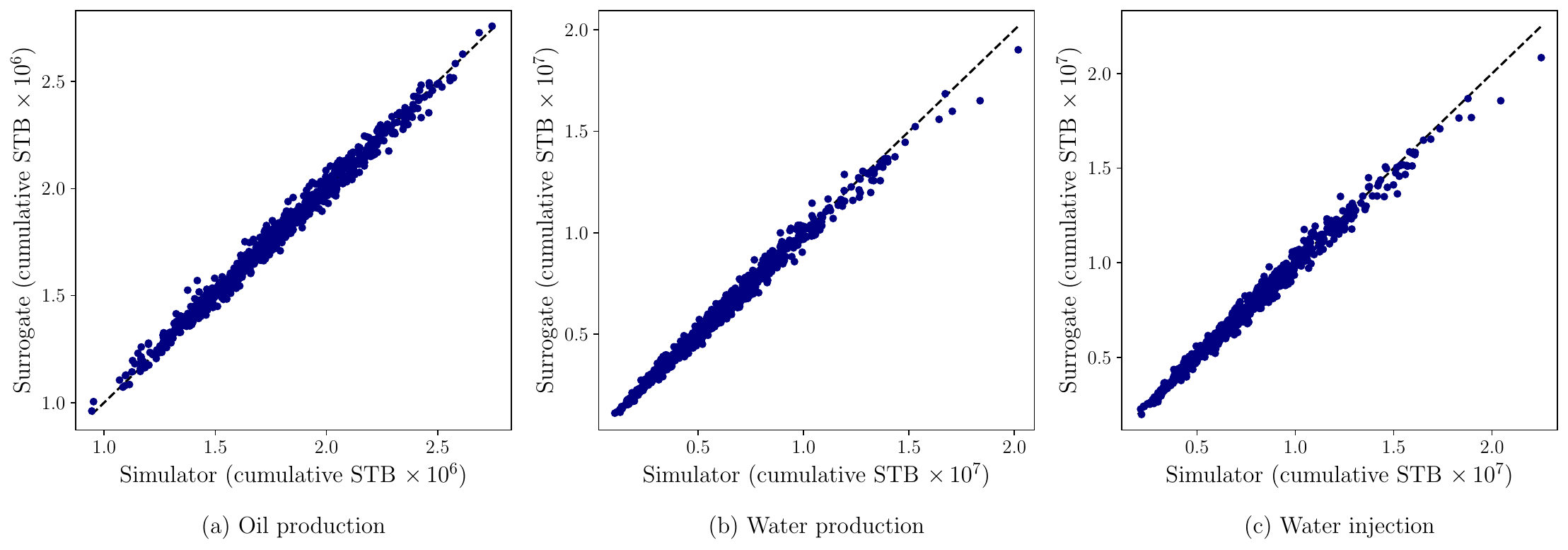}
\caption{Comparison of cumulative oil production, water production, and water injection predicted by the surrogate model and the high-fidelity simulator over the full 1500-day simulation period. Each point represents one test-set sample, and the black dashed line denotes the 1:1 reference.}
\label{fig:ow-surrogate-scatter}
\end{figure}

\begin{figure}[!htbp]
\centering
\includegraphics[width=0.95\linewidth]
{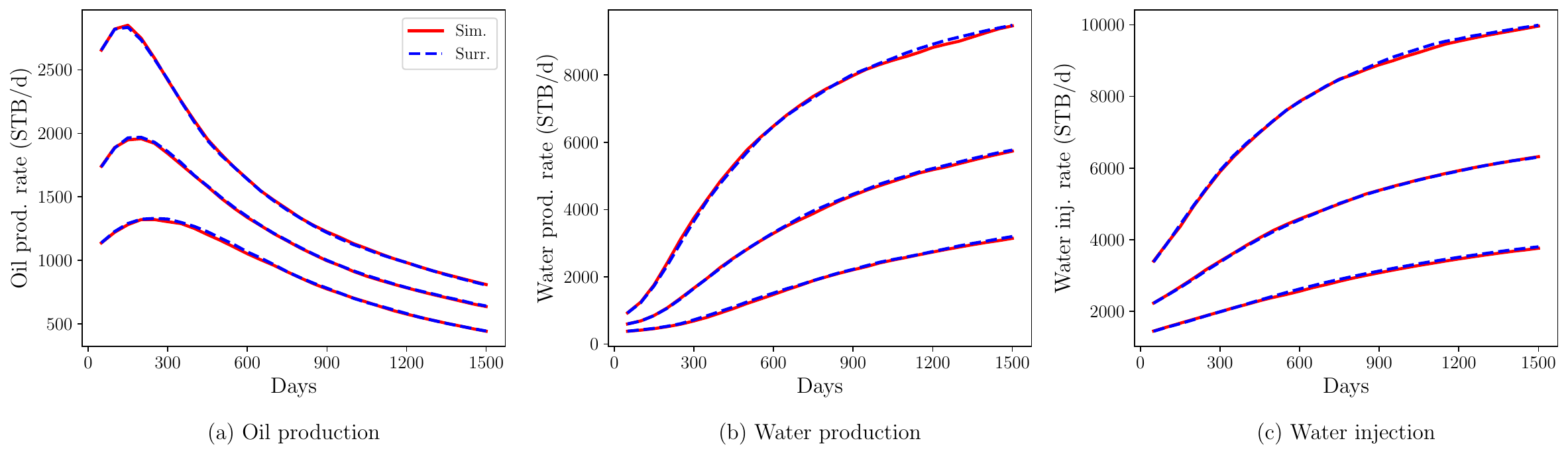}
\caption{Comparison of field-level flow rate statistics obtained from the high fidelity simulator (red) and the surrogate model (blue) over the test samples. The lower, middle, and upper dashed curves represent the $\mathrm{P}_{10}$, $\mathrm{P}_{50}$, and $\mathrm{P}_{90}$ responses, respectively.}
\label{fig:ow-surrogate-rates}
\end{figure}

Additional assessments of surrogate accuracy are provided in Figs.~\ref{fig:ow-surrogate-scatter} and~\ref{fig:ow-surrogate-rates}. In Fig.~\ref{fig:ow-surrogate-scatter} we compare cumulative field-level oil production, water production, and water injection from the surrogate and tNavigator for the test set. These cumulative quantities are integrals of the rates over the 1500-day simulation time frame (summed over all wells). The predictions closely follow the 1:1 reference, with $R^2=0.988$, $0.990$, and $0.989$, respectively. In Fig.~\ref{fig:ow-surrogate-rates} we present the $\mathrm{P}_{10}$, $\mathrm{P}_{50}$, and $\mathrm{P}_{90}$ (10th, 50th, 90th percentile) responses for field-level rates over the test set. Field-level rates are computed by summing over all wells. Very close agreement is observed in both the temporal evolution and the spread of these percentile curves. The consistency between the flow surrogate and the high-fidelity simulator, as demonstrated in Figs.~\ref{fig:ow-surrogate-scatter} and~\ref{fig:ow-surrogate-rates}, suggests that the flow surrogate is indeed applicable for use in data assimilation.

\FloatBarrier

\section{Sequential Monte Carlo for hierarchical data assimilation}
\label{sec:smc_hierarchical}

In this section, we first formulate the hierarchical data assimilation problem using the low-dimensional geological parameterizations introduced for the two subsurface flow systems. We then present the adaptive tempered sequential Monte Carlo (SMC) method for hierarchical data assimilation. Finally, the extension of SMC to successive data assimilation, in which observations become available at a series of times, is described.

\subsection{Hierarchical data assimilation}
\label{sec:hierarchical_da}

As discussed in Section~\ref{sec:sim_setup}, geological uncertainty is represented hierarchically in both subsurface systems. For the faulted carbon storage system, the geological metaparameters $\mathbf{h}$ are separated from the PCA latent variables $\boldsymbol{\xi}$, which describe realization-specific spatial variability. For the channelized system, the 3D-LDM represents variability in both the geometric metaparameters $\mathbf{h}_g$ and the corresponding facies realizations through $\boldsymbol{\xi}_T$. The petrophysical properties (metaparameters $\mathbf{h}_p$) are also uncertain in these models. To have common notation, we use $\mathbf{h}$ and $\boldsymbol{\xi}$ to denote the inferred metaparameter and latent vectors for either system. For the channelized system, these correspond to $\mathbf{h}_p$ and $\boldsymbol{\xi}_T$. Note that the term ``metaparameter'' is often used to describe higher-level properties (like $\mu_{\log k}$ and $\sigma_{\log k}$) that characterize the distribution. Here we use this term more generally to include both higher-level parameters along with any model parameter other than the latent variable $\boldsymbol{\xi}$.

The fully specified geomodel is denoted by $\mathbf{m}=\mathbf{m}(\mathbf{h},\boldsymbol{\xi})$. This representation corresponds to a noncentered parameterization, in which the latent variables $\boldsymbol{\xi}$ have a reference distribution that is independent of $\mathbf{h}$. The components of $\mathbf{h}$ can scale and shift the geomodel properties, but they do not change the spatial correlation structure.

We now describe the observed data and model error considered in our data assimilation procedures. Let $\mathbf{d}_{\mathrm{obs}}\in\mathbb{R}^{n_d}$ denote the observed data. In real field applications, $\mathbf{d}_{\mathrm{obs}}$ derive from observations at injection wells and monitoring wells (CO$_2$ storage) or injection and production wells (oil production). Here we consider synthetic data generated by the high-fidelity flow simulator. We denote the simulated data at the observation locations and times as $f(\mathbf{m})$. The observed data include noise and are expressed as
\begin{equation}
\label{noise}
\mathbf{d}_{\mathrm{obs}}
=
\mathbf{d}_{\mathrm{true}}+\boldsymbol{\epsilon}
=
f\!\left(\mathbf{m}_{\mathrm{true}}\right)
+\boldsymbol{\epsilon}.
\end{equation}
Here $\mathbf{m}_{\mathrm{true}}$ denotes the true geomodel and $\mathbf{d}_{\mathrm{true}}=f(\mathbf{m}_{\mathrm{true}})$ are the corresponding true data. Measurement errors are assumed to be independent and Gaussian, with $\boldsymbol{\epsilon}\sim\mathcal{N}(\mathbf{0},\mathbf{C}_d)$ and diagonal covariance matrix $\mathbf{C}_d\in\mathbb{R}^{n_d\times n_d}$.

During data assimilation, the high-fidelity simulator $f$ is replaced by the deep learning surrogate $\hat{f}$. Model error results from approximating $f(\mathbf{m})$ by $\hat{f}(\mathbf{m})$, and it is important to incorporate this error into the likelihood. Following~\citet{han2025accelerated}, surrogate error statistics are estimated from paired high-fidelity and surrogate predictions available from the $n_e$ test-set geomodels. For realization $j$, the error vector is
\begin{equation}
\label{eq:surrogate_error}
\mathbf{e}_{\mathrm{surr}}^{(j)}
=
f\!\left(\mathbf{m}^{(j)}\right)
-
\hat{f}\!\left(\mathbf{m}^{(j)}\right),
\qquad j=1,\ldots,n_e.
\end{equation}
The sample mean is $\bar{\mathbf{e}}_{\mathrm{surr}}=n_e^{-1}\sum_{j=1}^{n_e}\mathbf{e}_{\mathrm{surr}}^{(j)}$. Because our flow surrogates are essentially unbiased, this mean is close to zero. The surrogate error covariance, which is nonzero, is estimated as
\begin{equation}
\label{eq:surrogate_error_covariance}
\mathbf{C}_{\mathrm{surr}}
=
\frac{1}{n_e-1}
\sum_{j=1}^{n_e}
\left(
\mathbf{e}_{\mathrm{surr}}^{(j)}
-\bar{\mathbf{e}}_{\mathrm{surr}}
\right)
\left(
\mathbf{e}_{\mathrm{surr}}^{(j)}
-\bar{\mathbf{e}}_{\mathrm{surr}}
\right)^\top.
\end{equation}
Unlike the diagonal measurement-error covariance $\mathbf{C}_d$, $\mathbf{C}_{\mathrm{surr}}\in\mathbb{R}^{n_d\times n_d}$ is generally nondiagonal and captures correlations in surrogate errors across observation locations and times. The total error covariance used in the likelihood is
\begin{equation}
\label{eq:total_error_covariance}
\mathbf{C}_{\mathrm{tot}}
=
\mathbf{C}_d+\mathbf{C}_{\mathrm{surr}}.
\end{equation}
The likelihood, which now accounts for both measurement and surrogate model errors (through $\mathbf{C}_{\mathrm{tot}}$), is given by
\begin{equation}
\label{eq:likelihood_direct_31}
p(\mathbf{d}_{\mathrm{obs}}\mid\mathbf{h},\boldsymbol{\xi})
\propto
\exp\left[
-\frac{1}{2}
\left(
\mathbf{d}_{\mathrm{obs}}
-\hat{f}\!\left(\mathbf{m}(\mathbf{h},\boldsymbol{\xi})\right)
\right)^\top
\mathbf{C}_{\mathrm{tot}}^{-1}
\left(
\mathbf{d}_{\mathrm{obs}}
-\hat{f}\!\left(\mathbf{m}(\mathbf{h},\boldsymbol{\xi})\right)
\right)
\right].
\end{equation}

As  discussed earlier, under the noncentered parameterizations used here, the metaparameter and latent variables are independent under the prior. The joint prior is
\begin{equation}
\label{eq:prior_factorization}
p(\mathbf{h},\boldsymbol{\xi})
=
p(\mathbf{h})\,p(\boldsymbol{\xi}),
\qquad
\boldsymbol{\xi}
\sim
\mathcal{N}(\mathbf{0},\mathbf{I}_{n_\xi}),
\end{equation}
where $p(\mathbf{h})$ specifies the metaparameter priors given in Section~\ref{sec:sim_setup}, and $p(\boldsymbol{\xi})$ is the standard Gaussian density in the corresponding latent dimension. For the channelized system, $p(\mathbf{h})$ refers only to the petrophysical metaparameter prior, since geometric variability is already represented in the 3D-LDM. Bayes' rule then gives
\begin{equation}
\label{eq:posterior_pca_31}
p(\mathbf{h},\boldsymbol{\xi}\mid\mathbf{d}_{\mathrm{obs}})
\propto
p(\mathbf{d}_{\mathrm{obs}}\mid\mathbf{h},\boldsymbol{\xi})\,
p(\mathbf{h})\,p(\boldsymbol{\xi}).
\end{equation}
We now describe the use of SMC to provide samples from $p(\mathbf{h},\boldsymbol{\xi}\mid\mathbf{d}_{\mathrm{obs}})$.

\subsection{Adaptive tempered sequential Monte Carlo method}
\label{sec:smc}

\begin{figure}[!htbp]
\centering
\includegraphics[width=0.9\linewidth]{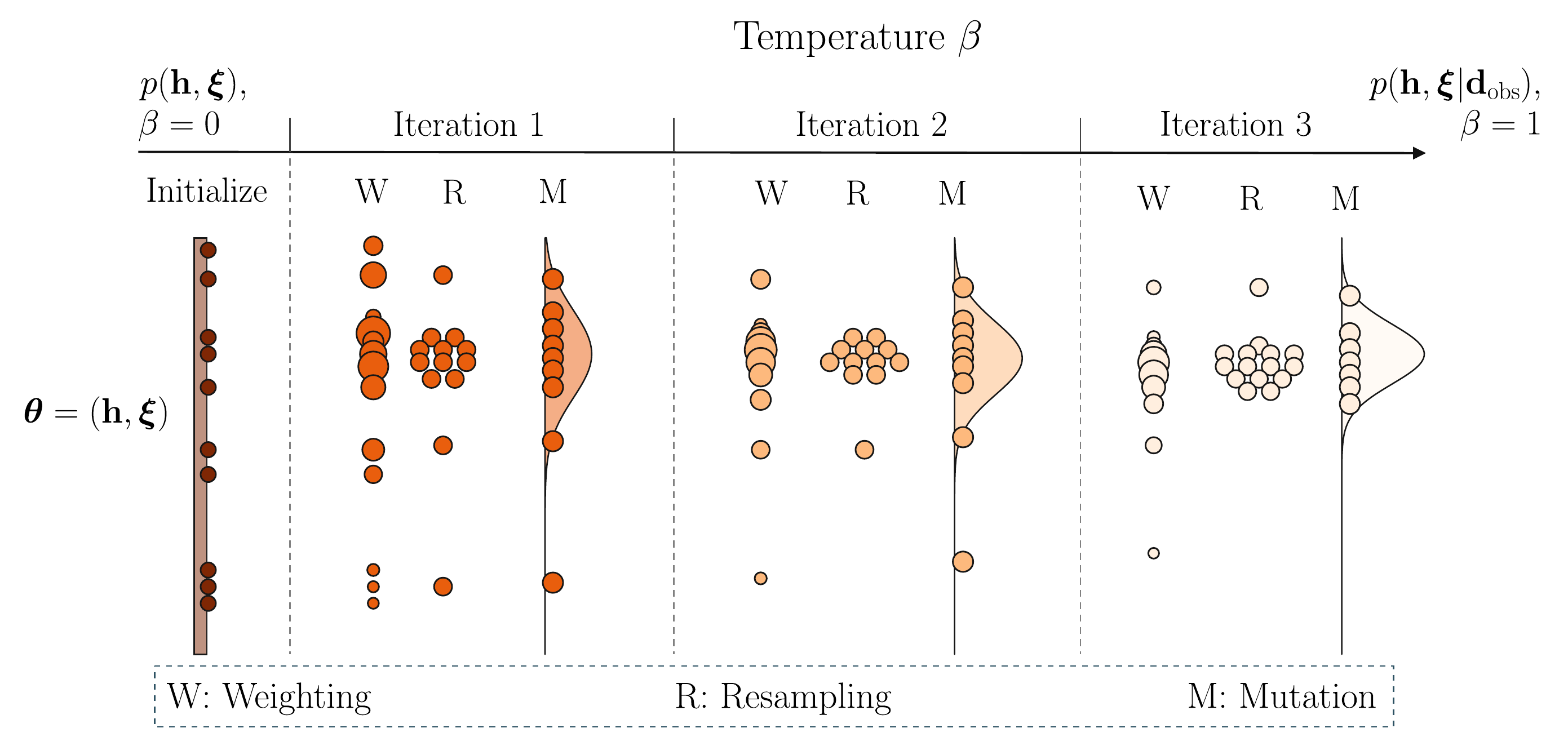}
\caption{Schematic illustration of the adaptive tempered sequential Monte Carlo algorithm over $k=3$ iterations (adapted from~\citep{Hinne2025}). Initial particles $\boldsymbol{\theta}=(\mathbf{h},\boldsymbol{\xi})$ are drawn from the prior distribution at $\beta=0$. At each iteration, particles are weighted by the likelihood, with relative weights indicated by marker size, resampled in proportion to their normalized weights, and then mutated via short MCMC chains targeting the corresponding tempered distribution. The procedure is repeated until $\beta$ reaches 1, where the particle population approximates the posterior distribution $p(\mathbf{h},\boldsymbol{\xi} \mid \mathbf{d}_{\mathrm{obs}})$.}
\label{fig:smc_sampler}
\end{figure}

Our adaptive tempered SMC sampler transports a population of particles, each representing a realization of $(\mathbf{h},\boldsymbol{\xi})$, from the joint prior $p(\mathbf{h},\boldsymbol{\xi})$ to the posterior in Eq.~\ref{eq:posterior_pca_31} through a sequence of intermediate tempered distributions. These intermediate distributions are indexed by the ``temperature'' parameter $\beta\in[0,1]$. As illustrated in Fig.~\ref{fig:smc_sampler}, the particles are initialized from the prior at $\beta=0$. At each subsequent iteration, they are weighted according to the likelihood, resampled using the normalized weights, and mutated using Markov moves that leave the current tempered distribution invariant. As $\beta$ increases to 1, the particle population is progressively transported toward the target posterior.

The sampler uses likelihood tempering to construct a sequence of intermediate target distributions between the prior and posterior. Let $0=\beta_0 < \beta_1 < \cdots < \beta_K=1$ denote the temperature schedule. The intermediate target densities are
\begin{equation}
    p_k(\mathbf{h},\boldsymbol{\xi}\mid \mathbf{d}_{\mathrm{obs}})
    \propto p(\mathbf{h})\,p(\boldsymbol{\xi})\,p(\mathbf{d}_{\mathrm{obs}} \mid \mathbf{h},\boldsymbol{\xi})^{\beta_k},
    \qquad k=0,\ldots,K.  
    \label{eq:tempered_targets}
\end{equation}
For $\beta_0=0$, the likelihood contribution is removed and the target reduces to the joint prior,
$p_0(\mathbf{h},\boldsymbol{\xi})=p(\mathbf{h})p(\boldsymbol{\xi})$, while for $\beta_K=1$, the full likelihood is recovered and the target becomes the posterior,
$p_K(\mathbf{h},\boldsymbol{\xi})
=p(\mathbf{h},\boldsymbol{\xi}\mid\mathbf{d}_{\mathrm{obs}})$.
The temperature $\beta_k$ therefore controls the contribution of the likelihood to the target distribution. By introducing the likelihood progressively, likelihood tempering reduces the particle weight degeneracy that can arise from direct importance weighting of prior samples by the full likelihood. This degeneracy can be particularly problematic in high-dimensional data assimilation problems.

The algorithm begins by drawing $N$ independent particles from the prior and assigning equal initial weights. It then advances from $p_{k-1}$ to $p_k$ through four steps: adaptive selection of the next temperature, importance weighting, resampling, and short-chain MCMC mutation. These steps are repeated until the terminal temperature $\beta_K=1$ is reached.

The temperature schedule can be prespecified using, e.g., linear or exponential spacing from $\beta=0$ to $\beta=1$. Adaptive tempering, which is applied here, is more robust because it limits particle weight degeneracy~\citep{Jasra2011}. We denote the equally weighted particles after the mutation step at iteration $k-1$ as
$(\mathbf{h}_{k-1}^{(i)},\boldsymbol{\xi}_{k-1}^{(i)})$, $i=1,\ldots,N$, and we assume they approximately follow the current tempered target $p_{k-1}$. For a candidate temperature increment
$\Delta\beta \in [0,1-\beta_{k-1}]$, the unnormalized importance weights are
\begin{equation}
    \widetilde{w}_k^{(i)}(\Delta\beta)
    =
    p(\mathbf{d}_{\mathrm{obs}} \mid \mathbf{h}_{k-1}^{(i)},\boldsymbol{\xi}_{k-1}^{(i)})^{\Delta\beta},
    \qquad i=1,\ldots,N .
    \label{eq:candidate_incremental_weights_section32}
\end{equation}
These weights account for the additional likelihood contribution introduced when the temperature increases from $\beta_{k-1}$ to $\beta_{k-1}+\Delta\beta$.

The temperature increment is determined by the effective sample size (ESS) associated with the importance weights~\cite{Kong1994},
\begin{equation}
    \operatorname{ESS}(\Delta\beta)
    =
    \frac{
    \left(\sum_{i=1}^N \widetilde{w}_k^{(i)}(\Delta\beta)\right)^2
    }{
    \sum_{i=1}^N \left[\widetilde{w}_k^{(i)}(\Delta\beta)\right]^2
    }
    =
    \frac{
    \left(\sum_{i=1}^N
    p(\mathbf{d}_{\mathrm{obs}} \mid \mathbf{h}_{k-1}^{(i)},\boldsymbol{\xi}_{k-1}^{(i)})^{\Delta\beta}
    \right)^2
    }{
    \sum_{i=1}^N
    p(\mathbf{d}_{\mathrm{obs}} \mid \mathbf{h}_{k-1}^{(i)},\boldsymbol{\xi}_{k-1}^{(i)})^{2\Delta\beta}
    } .
    \label{eq:ess_section32}
\end{equation}
Here the numerator corresponds to the square of the total unnormalized weights and the denominator is the sum of the squared weights. Their ratio quantifies the concentration of the weights across particles. Thus, the ESS is close to $N$ when the weights are nearly uniform, and it decreases toward one when the weight is concentrated in only a few particles. For a specified ESS fraction $\rho \in (0,1)$, the algorithm chooses the largest temperature increment that maintains the target particle diversity level. If the remaining step to the posterior satisfies
$\operatorname{ESS}(1-\beta_{k-1}) \geq \rho N$, then the sampler advances to $\beta=1$ by setting
$\Delta\beta_k = 1-\beta_{k-1}$. Otherwise, $\Delta\beta_k$ is obtained by solving
\begin{equation}
    \operatorname{ESS}(\Delta\beta_k)=\rho N,
    \qquad
    \Delta\beta_k\in(0,1-\beta_{k-1}),
    \label{eq:ess_root_section32}
\end{equation}
using a numerical root-finding method. The temperature is then updated via $\beta_k=\beta_{k-1}+\Delta\beta_k$. This adaptive temperature selection is used in all numerical examples considered in this work. In our implementation we specify $\rho=0.5$.

The second step is importance weighting. After the adaptive increment $\Delta\beta_k$ has been selected, each particle is assigned the unnormalized weight
$\widetilde{w}_k^{(i)}=\widetilde{w}_k^{(i)}(\Delta\beta_k)$, as defined in Eq.~\ref{eq:candidate_incremental_weights_section32}. The normalized weights are then
\begin{equation}
    w_k^{(i)}
    =
    \frac{\widetilde{w}_k^{(i)}}{\sum_{j=1}^N\widetilde{w}_k^{(j)}} ,
    \qquad i=1,\ldots,N.
    \label{eq:normalized_weights_section32}
\end{equation}
The resulting weighted particle system provides an importance sampling approximation of the current tempered target.

The third step is resampling, in which the weighted ensemble is replaced by an ensemble with equally weighted particles. This is accomplished by drawing particles according to the normalized weights $w_k^{(i)}$. Thus, particles with large weights may be replicated, whereas particles with negligible weights may be removed. Several resampling schemes are commonly used in SMC methods, including multinomial, residual, stratified, and systematic resampling. In this work, we adopt systematic resampling~\cite{Kitagawa1996,Carpenter1999}. Because resampling can reduce particle diversity through duplication, it is followed by MCMC mutation steps that move the particles while preserving the current tempered distribution $p_k(\mathbf{h},\boldsymbol{\xi}\mid \mathbf{d}_{\mathrm{obs}})$.

The mutation step applied after resampling entails running a very short Markov chain of $M$ steps for each resampled particle, with $M$ typically set to $10$--$60$. Unlike a standalone MCMC sampler, which generally requires sufficiently long chains to obtain reliable posterior estimates, the short chains used here serve only for particle mutation within the SMC procedure. The mutation step applies Markov transition kernels that leave the current tempered distribution invariant. Each MCMC step is implemented as a blockwise Metropolis-within-Gibbs update. For a resampled particle, let
$(\mathbf{h}^{\ell-1},\boldsymbol{\xi}^{\ell-1})$ denote the current state. The high-dimensional latent variable block is updated first using the preconditioned Crank--Nicolson (pCN) proposal~\cite{Cotter2013}
\begin{equation}
\boldsymbol{\xi}'
=
\left(1-s_{\xi}^2\right)^{\frac {1}{2}}\,\boldsymbol{\xi}^{\ell-1}
+s_{\xi}\boldsymbol{\eta},
\qquad
\boldsymbol{\eta}\sim \mathcal{N}(\mathbf{0},\mathbf{I}_{n_\xi}),
\label{eq:smc_pcn_proposal_refined}
\end{equation}
where $s_{\xi}\in(0,1)$ is the pCN proposal scale. Because this proposal is reversible with respect to the standard Gaussian prior, the Gaussian prior and proposal-density terms cancel in the Metropolis--Hastings ratio. The proposed latent variable is accepted with probability
\begin{equation}
\alpha_{\xi}
\left(\boldsymbol{\xi}^{\ell-1},\boldsymbol{\xi}'\right)
=
\min\left\{1,
\left[
\frac{
p(\mathbf{d}_{\mathrm{obs}}\mid \mathbf{h}^{\ell-1},\boldsymbol{\xi}')
}{
p(\mathbf{d}_{\mathrm{obs}}\mid \mathbf{h}^{\ell-1},\boldsymbol{\xi}^{\ell-1})
}
\right]^{\beta_k}
\right\}.
\label{eq:smc_pcn_acceptance_refined}
\end{equation}
Thus, after drawing $u_\xi\sim U(0,1)$, we set
$\boldsymbol{\xi}^{\ell}=\boldsymbol{\xi}'$ if $u_\xi<\alpha_\xi$; otherwise,
$\boldsymbol{\xi}^{\ell}=\boldsymbol{\xi}^{\ell-1}$.

The low-dimensional metaparameter block is then updated conditional on the current latent variable $\boldsymbol{\xi}^{\ell}$. We use a Gaussian random walk proposal,
\begin{equation}
\mathbf{h}'
=
\mathbf{h}^{\ell-1}+\boldsymbol{\zeta}_{h},
\qquad
\boldsymbol{\zeta}_{h}\sim \mathcal{N}(\mathbf{0},\Sigma_{h}),
\label{eq:smc_h_rw_proposal_refined}
\end{equation}
where $\Sigma_{h}$ is the proposal covariance. The proposal density terms again cancel in the Metropolis--Hastings ratio. The proposed metaparameter vector is accepted with probability
\begin{equation}
\alpha_h
\left(\mathbf{h}^{\ell-1},\mathbf{h}'\right)
=
\min\left\{1,
\frac{p_{\mathbf{h}}(\mathbf{h}')}{p_{\mathbf{h}}(\mathbf{h}^{\ell-1})}
\left[
\frac{
p(\mathbf{d}_{\mathrm{obs}}\mid \mathbf{h}',\boldsymbol{\xi}^{\ell})
}{
p(\mathbf{d}_{\mathrm{obs}}\mid \mathbf{h}^{\ell-1},\boldsymbol{\xi}^{\ell})
}
\right]^{\beta_k}
\right\}.
\label{eq:smc_h_acceptance_refined}
\end{equation}
After drawing $u_h\sim U(0,1)$, we set
$\mathbf{h}^{\ell}=\mathbf{h}'$ if $u_h<\alpha_h$ and
$\mathbf{h}^{\ell}=\mathbf{h}^{\ell-1}$ otherwise.

The proposal scales are tuned to provide effective particle mutation while maintaining reasonable acceptance probabilities. Specifically, the pCN proposal scale $s_{\xi}$ in Eq.~\ref{eq:smc_pcn_proposal_refined} and the scale of the Gaussian random-walk covariance $\Sigma_h$ in Eq.~\ref{eq:smc_h_rw_proposal_refined} are tuned such that the corresponding acceptance probabilities in Eqs.~\ref{eq:smc_pcn_acceptance_refined} and~\ref{eq:smc_h_acceptance_refined}, averaged over the particle population, are approximately $10\%$--$30\%$ at the final iteration.

The Markov updates applied here are used only to move the resampled particles while preserving the current tempered distribution. They should not be viewed as independent posterior chains, and no separate burn-in period is introduced. Because the mutation kernels are designed to leave $p_k$ invariant, the equally weighted ensemble after mutation remains an approximation of the current tempered target. The details of the adaptive tempered SMC algorithm for hierarchical data assimilation, used for both the faulted carbon storage and channelized oil-water systems, are provided in Algorithm~\ref{alg:adaptive_tempered_smc}.

Compared with a standard MCMC sampler, our adaptive tempered SMC algorithm provides a population-based posterior approximation without a separate burn-in stage. The population-based representation acts to maintain samples in multiple regions of the target posterior, which is useful for multimodal Bayesian estimation problems. Unlike standard MCMC, the SMC algorithm parallelizes naturally because likelihood evaluations and mutation updates can be performed independently for each particle. In addition, its structure is well suited to successive data assimilation, where observations become available at a series of times. In this case the existing particle population can be updated rather than restarting the full inference using all available observations. This extension of the adaptive tempered SMC procedure for successive data assimilation is described next.

\begin{algorithm}[!htbp]
\caption{Adaptive tempered SMC for hierarchical data assimilation}
\label{alg:adaptive_tempered_smc}
\begin{algorithmic}[1]
\State Set $k=0$, $\beta_0=0$, the number of particles $N$, ESS fraction $\rho$, mutation length $M$, pCN scale $s_\xi$, and Gaussian random-walk proposal covariance $\Sigma_h$. 
\For{$i=1,\ldots,N$}
    \State Draw $\mathbf{h}_0^{(i)}\sim p(\mathbf{h})$ and $\boldsymbol{\xi}_0^{(i)}\sim \mathcal{N}(\mathbf{0},\mathbf{I}_{n_\xi})$.
    \State Set $w_0^{(i)}=1/N$.
\EndFor

\While{$\beta_k<1$}
    \State Set $k \leftarrow k+1$.
    \If{$\operatorname{ESS}(1-\beta_{k-1})\geq \rho N$}
        \State Set $\beta_{k}=1$.
    \Else
        \State Solve Eq.~\eqref{eq:ess_root_section32} to obtain $\Delta\beta_{k}$.
        \State Set $\beta_{k}=\beta_{k-1}+\Delta\beta_{k}$.
    \EndIf

    \State Compute unnormalized importance weights using Eq.~\eqref{eq:candidate_incremental_weights_section32}.
    \State Normalize the weights using Eq.~\eqref{eq:normalized_weights_section32}.
    \If{$\beta_k=1$}
        \State Set $(\mathbf{h}_{k}^{(i)},\boldsymbol{\xi}_{k}^{(i)})=(\mathbf{h}_{k-1}^{(i)},\boldsymbol{\xi}_{k-1}^{(i)})$, $w_k^{(i)}=w_{k}^{(i)}$
    \Else
        \State Resample $\{(\mathbf{h}_{k}^{(i)},\boldsymbol{\xi}_{k}^{(i)})\}_{i=1}^N$ using $\{w_{k}^{(i)}\}_{i=1}^N$, and set $w_k^{(i)}=1/N$.
    
        \For{$i=1,\ldots,N$}
            \For{$m=1,\ldots,M$}
                \State Propose $\boldsymbol{\xi}'$ using the pCN proposal in Eq.~\eqref{eq:smc_pcn_proposal_refined}.
                \State Accept or reject $\boldsymbol{\xi}'$ using Eq.~\eqref{eq:smc_pcn_acceptance_refined}.
                \State Propose $\mathbf{h}'$ using the Gaussian random-walk proposal in Eq.~\eqref{eq:smc_h_rw_proposal_refined}.
                \State Accept or reject $\mathbf{h}'$ using Eq.~\eqref{eq:smc_h_acceptance_refined}.
            \EndFor
            \State Set $(\mathbf{h}_{k}^{(i)},\boldsymbol{\xi}_{k}^{(i)})=(\mathbf{h},\boldsymbol{\xi})$.
        \EndFor
    \EndIf
        
\EndWhile

\end{algorithmic}
\end{algorithm}

\subsection{Successive data assimilation}
\label{sec:sequential_da}

Let $\mathbf{d}_{\mathrm{obs},1}\in\mathbb{R}^{n_{d,1}}$ denote the previously assimilated observations and $\mathbf{d}_{\mathrm{obs},2}\in\mathbb{R}^{n_{d,2}}$ a new batch of observations. The posterior after assimilating $\mathbf{d}_{\mathrm{obs},2}$ can be expressed as
\begin{equation}
\label{eq:sequential_posterior}
p(\boldsymbol{\theta}
\mid\mathbf{d}_{\mathrm{obs},1},\mathbf{d}_{\mathrm{obs},2})
\propto
p(\boldsymbol{\theta}\mid\mathbf{d}_{\mathrm{obs},1})
\,p(\mathbf{d}_{\mathrm{obs},2}
\mid\mathbf{d}_{\mathrm{obs},1},\boldsymbol{\theta}),
\end{equation}
where $\boldsymbol{\theta}=(\mathbf{h},\boldsymbol{\xi})$. The second factor is the conditional likelihood of the new observations. This form is used because the surrogate model error included in the total error covariance can introduce correlations between the two batches of observations. These correlations are therefore retained explicitly in the successive estimation procedure (they would be neglected under a conditional-independence assumption).

For the Gaussian total error model introduced in Section~\ref{sec:hierarchical_da}, the joint likelihood for the two observation batches is
\begin{equation}
\label{eq:sequential_joint_likelihood}
\begin{aligned}
&p(\mathbf{d}_{\mathrm{obs},1},\mathbf{d}_{\mathrm{obs},2}
\mid\boldsymbol{\theta})
=
\mathcal{N}\left(
\begin{bmatrix}
\mathbf{d}_{\mathrm{obs},1}\\
\mathbf{d}_{\mathrm{obs},2}
\end{bmatrix};
\begin{bmatrix}
\hat{f}_1(\boldsymbol{\theta})\\
\hat{f}_2(\boldsymbol{\theta})
\end{bmatrix},
\begin{bmatrix}
\mathbf{C}_{11} & \mathbf{C}_{12}\\
\mathbf{C}_{21} & \mathbf{C}_{22}
\end{bmatrix}
\right),
\end{aligned}
\end{equation}
where $\mathcal{N}(\cdot;\boldsymbol{\mu},\mathbf{C})$ denotes a Gaussian density with mean $\boldsymbol{\mu}$ and covariance $\mathbf{C}$. Here $\hat{f}_j(\boldsymbol{\theta})$, $j=1,2$, denotes the surrogate predictions for observation batch $j$, and $\mathbf{C}_{ij}$ are the corresponding blocks of the total error covariance $\mathbf{C}_{\mathrm{tot}}$, with $\mathbf{C}_{21}=\mathbf{C}_{12}^{\top}$. These covariance blocks are fixed during inference, and $\mathbf{C}_{11}$ is the covariance associated with the first observation batch.

To simplify the notation for successive data assimilation, let $L_1(\boldsymbol{\theta})=p(\mathbf{d}_{\mathrm{obs},1}\mid\boldsymbol{\theta})$ denote the likelihood of the previously assimilated observations, and let $L_2(\boldsymbol{\theta})=p(\mathbf{d}_{\mathrm{obs},2}\mid\mathbf{d}_{\mathrm{obs},1},\boldsymbol{\theta})$ denote the conditional likelihood of the additional observations. Because the conditional distribution of a jointly Gaussian vector is also Gaussian, Eq.~\ref{eq:sequential_joint_likelihood} gives
\begin{equation}
\label{eq:sequential_conditional_distribution}
\begin{aligned}
L_2(\boldsymbol{\theta})
&=
\mathcal{N}\!\left(
\mathbf{d}_{\mathrm{obs},2};
\boldsymbol{\mu}_{2\mid1}(\boldsymbol{\theta}),
\mathbf{C}_{2\mid1}
\right),
\end{aligned}
\end{equation}
where the conditional mean and covariance are
\begin{equation}
\label{eq:sequential_conditional_moments}
\begin{aligned}
\boldsymbol{\mu}_{2\mid1}(\boldsymbol{\theta})
&=
\hat{f}_2(\boldsymbol{\theta})
+
\mathbf{C}_{21}\mathbf{C}_{11}^{-1}
\left[
\mathbf{d}_{\mathrm{obs},1}
-\hat{f}_1(\boldsymbol{\theta})
\right],\\
\mathbf{C}_{2\mid1}
&=
\mathbf{C}_{22}
-
\mathbf{C}_{21}\mathbf{C}_{11}^{-1}\mathbf{C}_{12}.
\end{aligned}
\end{equation}
Using these conditional moments, the conditional likelihood for the second observation batch can be written as
\begin{equation}
\label{eq:sequential_conditional_likelihood}
L_2(\boldsymbol{\theta})
\propto
\exp\left[
-\frac{1}{2}
\left(
\mathbf{d}_{\mathrm{obs},2}
-\boldsymbol{\mu}_{2\mid1}(\boldsymbol{\theta})
\right)^\top
\mathbf{C}_{2\mid1}^{-1}
\left(
\mathbf{d}_{\mathrm{obs},2}
-\boldsymbol{\mu}_{2\mid1}(\boldsymbol{\theta})
\right)
\right].
\end{equation}
The cross-covariance term therefore accounts for correlations between the errors in the two observation batches. In the absence of cross-batch error correlations, $\mathbf{C}_{12}=\mathbf{0}$ (and hence $\mathbf{C}_{21}=\mathbf{0}$), the conditional mean and covariance reduce to $\hat{f}_2(\boldsymbol{\theta})$ and $\mathbf{C}_{22}$, respectively, and $L_2$ becomes the marginal likelihood of the second observation batch.

The conditional likelihood $L_2$ is incorporated through a new adaptive tempering sequence,
$0=\beta_0<\beta_1<\cdots<\beta_K=1$, where $k$ indexes the tempering iterations used to assimilate the additional observations. The intermediate target densities are now
\begin{equation}
\label{eq:sequential_target}
p_k(\boldsymbol{\theta}
\mid\mathbf{d}_{\mathrm{obs},1},\mathbf{d}_{\mathrm{obs},2})
\propto
p(\mathbf{h})\,p(\boldsymbol{\xi})\,
L_1(\boldsymbol{\theta})\,
L_2(\boldsymbol{\theta})^{\beta_k},
\qquad k=0,\ldots,K.
\end{equation}
At $\beta_0=0$, the target is the posterior conditioned on the first observation batch. As $\beta_k$ increases, the contribution of the conditional likelihood for the second batch is introduced progressively. At $\beta_K=1$, the full conditional likelihood is recovered and the target becomes the posterior conditioned on both observation batches. Throughout this update, $L_1$ remains fully incorporated, while only $L_2$ is tempered.

The successive update is initialized using the weighted particles from the first assimilation stage. At $\beta_0=0$, the particles are resampled according to these weights and assigned equal weights. Blockwise MCMC mutation is then applied with the posterior conditioned on the first observation batch as the invariant target. The resulting equally weighted particles provide the initial population for tempering the conditional likelihood $L_2$.

Let $\boldsymbol{\theta}_{k-1}^{(i)}$, $i=1,\ldots,N$, denote the equally weighted particles after mutation at iteration $k-1$. For a candidate temperature increment
$\Delta\beta\in[0,1-\beta_{k-1}]$,
the unnormalized importance weights are
\begin{equation}
\label{eq:sequential_weights}
\widetilde{w}_k^{(i)}(\Delta\beta)
=
L_2\!\left(\boldsymbol{\theta}_{k-1}^{(i)}\right)^{\Delta\beta},
\qquad i=1,\ldots,N.
\end{equation}
Only $L_2$ appears in the incremental weights because the prior and the first-batch likelihood are unchanged between sequential tempered targets. The temperature increment is selected using the ESS criterion in Eqs.~\ref{eq:ess_section32} and~\ref{eq:ess_root_section32}, with the weights in Eq.~\ref{eq:sequential_weights} and the same threshold $\rho N$. The temperature is then updated as $\beta_k=\beta_{k-1}+\Delta\beta_k$, and the corresponding importance weights are normalized using Eq.~\ref{eq:normalized_weights_section32}.

At each intermediate tempering iteration, systematic resampling is followed by blockwise MCMC mutation targeting the current tempered distribution in Eq.~\ref{eq:sequential_target}. The pCN and Gaussian random-walk proposals remain unchanged, while the likelihood contribution to the corresponding acceptance probabilities is
\begin{equation}
\label{eq:successive_ratio}
R_k(\boldsymbol{\theta},\boldsymbol{\theta}')
=
\frac{L_1(\boldsymbol{\theta}')}{L_1(\boldsymbol{\theta})}
\left[
\frac{L_2(\boldsymbol{\theta}')}{L_2(\boldsymbol{\theta})}
\right]^{\beta_k},
\end{equation}
where $\boldsymbol{\theta}$ and $\boldsymbol{\theta}'$ denote the current and proposed states and differ only in the parameter block being updated. For the pCN update, $R_k$ replaces the tempered likelihood ratio in Eq.~\ref{eq:smc_pcn_acceptance_refined}. For the metaparameter update, $R_k$ replaces the corresponding likelihood ratio in Eq.~\ref{eq:smc_h_acceptance_refined}, while the metaparameter prior ratio is retained. The factor $L_1(\boldsymbol{\theta}')/L_1(\boldsymbol{\theta})$ is required because the MCMC mutation kernels must leave the full tempered target in Eq.~\ref{eq:sequential_target} invariant. Thus, the mutation step must account for the likelihood contributions from both observation batches, whereas the incremental importance weights depend only on $L_2$. Because the surrogate model provides fast predictions over the full simulation period, these likelihood contributions can be evaluated without separate surrogate forward evaluations for the two observation batches. The computational cost of the MCMC mutation step may increase, however, if longer mutation chains are needed to maintain sufficient particle diversity. At $\beta_K=1$, the resulting weighted particle population approximates the posterior conditioned on both observation batches.

The procedure can be extended to additional batches of observations in the same manner. For each new batch, the weighted posterior particles from the preceding assimilation stage are used to initialize a new adaptive tempering sequence, with the conditional likelihood of the new observations given all previously assimilated data incorporated progressively. This treatment retains correlations in surrogate model errors across observation batches and enables successive data assimilation as additional observations become available.

\section{Markov chain Monte Carlo method}
\label{sec:mcmc}

We use an existing Markov chain Monte Carlo (MCMC) procedure to obtain reference posterior results for assessing the adaptive tempered SMC method for hierarchical data assimilation. The MCMC calculations target the joint posterior $p(\mathbf{h},\boldsymbol{\xi}\mid\mathbf{d}_{\mathrm{obs}})$ in Eq.~\ref{eq:posterior_pca_31}, with the surrogate-based likelihood in Eq.~\ref{eq:likelihood_direct_31}, as described in Section~\ref{sec:hierarchical_da}. The likelihood accounts for measurement error and correlated surrogate model error through $\mathbf{C}_{\mathrm{tot}}$. In contrast to the short-chain MCMC mutation step used in SMC, the reference MCMC uses multiple independent Markov chains that target the posterior distribution, without likelihood tempering, importance weighting, or resampling.

For each of the two subsurface systems considered in this study, we run five independent Markov chains. Each chain is initialized from the joint prior, with the metaparameters sampled from $p(\mathbf{h})$ and the latent variables sampled from $\mathcal{N}(\mathbf{0},\mathbf{I}_{n_\xi})$. At each iteration, the latent variables are updated first using the pCN proposal in Eq.~\ref{eq:smc_pcn_proposal_refined}. The corresponding Metropolis--Hastings acceptance probability is given by Eq.~\ref{eq:smc_pcn_acceptance_refined}. Because the MCMC chains target the full posterior, the acceptance probability is
\begin{equation}
\label{eq:mcmc_pcn_acceptance}
\alpha_{\xi}
=
\min\left\{
1,
\frac{
p(\mathbf{d}_{\mathrm{obs}}\mid\mathbf{h}^{\ell-1},\boldsymbol{\xi}')
}{
p(\mathbf{d}_{\mathrm{obs}}\mid\mathbf{h}^{\ell-1},\boldsymbol{\xi}^{\ell-1})
}
\right\}.
\end{equation}
The metaparameters are then updated conditional on the accepted or retained latent state $\boldsymbol{\xi}^{\ell}$, using the Gaussian random-walk proposal in Eq.~\ref{eq:smc_h_rw_proposal_refined}. The corresponding acceptance probability is given by Eq.~\ref{eq:smc_h_acceptance_refined} with $\beta_k=1$,
\begin{equation}
\label{eq:mcmc_h_acceptance}
\alpha_h
=
\min\left\{
1,
\frac{p_{\mathbf{h}}(\mathbf{h}')}{p_{\mathbf{h}}(\mathbf{h}^{\ell-1})}
\frac{
p(\mathbf{d}_{\mathrm{obs}}\mid\mathbf{h}',\boldsymbol{\xi}^{\ell})
}{
p(\mathbf{d}_{\mathrm{obs}}\mid\mathbf{h}^{\ell-1},\boldsymbol{\xi}^{\ell})
}
\right\}.
\end{equation}
These two block updates form one MCMC iteration.

The first 5000 iterations of each chain are discarded as burn-in. Because we are interested in achieving (essentially) converged MCMC solutions, we apply strict convergence criteria. Specifically, convergence of the metaparameters is assessed using the rank-normalized and folded split-$\widehat{R}$ diagnostic of~\citet{Vehtari2021}. In the split-chain procedure, the post-burn-in samples from each of the five chains are divided into two halves. The resulting ten half-chains are treated as separate chains to improve the detection of nonstationarity within individual chains. For each metaparameter $h_j$, the reported value $\widehat{R}_j$ is the larger of the rank-normalized split-$\widehat{R}$ and the corresponding folded diagnostic, thereby accounting for differences in both location and scale among the chains. Values of $\widehat{R}_j$ approaching one indicate increasing agreement among the chains. In this study, the MCMC calculation is considered converged when
\begin{equation}
\label{eq:mcmc_convergence}
\max_{j=1,\ldots,n_h}\widehat{R}_j < 1.05,
\end{equation}
where $n_h$ is the number of geological metaparameters. The MCMC results at the first point satisfying Eq.~\ref{eq:mcmc_convergence} provide a set of reference posterior results for comparison with SMC.

Because this threshold is less stringent than the value of 1.01 recommended by~\citet{Vehtari2021}, we also continue the MCMC sampling to evaluate the sensitivity of posterior results to additional chain evolution. Our strategy for this is as follows. After Eq.~\ref{eq:mcmc_convergence} is first satisfied, the same Markov chains are continued until the stopping rule
\begin{equation}
\label{eq:mcmc_stopping_rule}
\max_{j=1,\ldots,n_h}\widehat{R}_j < 1.01
\quad \text{or} \quad
T \geq T_{\max},
\end{equation}
is met. Here $T$ is the total elapsed time of the MCMC calculation, and we set $T_{\max}=50~\mathrm{hours}$. This time limit provides a practical bound on this additional computation and does not imply satisfaction of the more stringent $\widehat{R}$ condition. We will present metaparameter posterior distributions obtained at the convergence point defined by Eq.~\ref{eq:mcmc_convergence} along with those obtained at the stopping point in Eq.~\ref{eq:mcmc_stopping_rule}. This comparison provides a direct assessment of whether the $\widehat{R}<1.05$ criterion yields posterior estimates that remain stable under additional MCMC sampling. Both sets of MCMC results will be compared with those from SMC.

As an additional convergence diagnostic, we compute the bulk effective sample size (bulk ESS) for each geological metaparameter as in~\citep{Vehtari2021}. The bulk ESS is based on rank-normalized split-chain samples and accounts for serial correlation within the Markov chains, thereby providing a measure of the effective number of samples available to characterize the central region of the posterior distribution. Larger bulk ESS values indicate more efficient exploration of this region. Bulk ESS is used here as a supplementary diagnostic and does not enter either the convergence criterion in Eq.~\ref{eq:mcmc_convergence} or the stopping rule in Eq.~\ref{eq:mcmc_stopping_rule}. We note that this diagnostic differs from the importance-weight ESS in Eq.~\ref{eq:ess_section32}, which is used to determine the adaptive tempering schedule in SMC. More specifically, the importance-weight ESS measures the concentration of particle weights, but it does not account for dependence among samples (and equally weighted samples are not necessarily independent). If applied to MCMC samples, which are equally weighted but serially correlated, this measure would simply return the number of retained samples, whereas the bulk ESS accounts for this correlation.

MCMC is also used to provide posterior results for successive data assimilation. Conventional MCMC is not naturally formulated for successive Bayesian estimation and generally requires a new sampling run as new observations become available~\citep{Berzuini1997,DelMoral2006,kantas2014sequential}. In our implementation, the new MCMC run is initialized from the joint prior. This MCMC computation targets the posterior conditioned on all observations available up to that stage (with all error correlations incorporated in $\mathbf{C}_{\mathrm{tot}}$). The successive SMC computations, by contrast, start from the posterior particle population obtained at the previous assimilation stage, as explained in Section~\ref{sec:sequential_da}. The SMC and MCMC approaches target the same updated posterior distribution, but reach it through different sampling procedures. The SMC approach is expected to be much more efficient in this setting, since it makes use of data assimilation results from previous stages.

\section{Data assimilation results for the faulted carbon storage system}
\label{sec:carbon_results}

In this section, we apply the SMC procedure for hierarchical data assimilation in the faulted carbon storage system described in Section~\ref{sec:co2_geomodel}. The SMC posterior estimates are compared with reference results obtained using MCMC. The results are evaluated in terms of posterior metaparameter distributions, CO$_2$ plume forecasts, saturation and pressure predictions at monitoring wells, and computational efficiency. In our comparisons in this paper, we consider only formal sampling procedures, not ensemble Kalman methods. This is because \citet{difederico2026data} showed that ensemble Kalman methods, when used in combination with highly nonlinear LDM parameterizations, can substantially overestimate posterior uncertainty. These results were obtained for channelized geomodels like those considered here, but without hierarchical uncertainty of the type treated in this study.

For Case~1, we consider observations through year~6 (Case~1A) and through year~12 (Case~1B), with the latter used to evaluate successive data assimilation. Detailed results for a second true model (Case~3) are provided in Supplementary Material (SM), though computational efficiency results for Cases~1 and~3 are provided in this section. The observed data comprise CO$_2$ saturation and pressure measurements at the three monitoring wells in all 37~layers of the model. This includes 25~layers in the target aquifer and 6~layers in each of the middle and upper aquifers. In Case~1A, observations are collected at 2, 4, and 6~years after the start of injection. This gives $3\times 37\times 3=333$ measurements each for saturation and pressure. We consider synthetic observed data generated using simulated data for the true model combined with Gaussian measurement noise. The standard deviation of measurement error is 0.02 saturation units for saturation and 0.1~MPa for pressure.

In Case~1B, we consider successive data assimilation using SMC, initialized from the posterior particles obtained at year~6 in Case~1A. These particles are updated using additional observations at 8, 10, and 12~years. The second batch adds 666 measurements, leading to a total of 1332 measurements through year~12. By contrast, the MCMC procedure requires a new run initialized from the prior using all 1332 observations through year~12. For comparison, a separate single-stage SMC calculation is also initialized from the prior using the same observations.

The MCMC calculations follow the procedure described in Section~\ref{sec:mcmc}, with five independent chains and the first 5000 iterations of each chain discarded as burn-in. All SMC calculations use $N=1000$ particles and a fixed importance-weight ESS threshold fraction of $\rho=0.5$. Each MCMC and SMC calculation uses a single NVIDIA A100 GPU, with 40~GB of memory for Cases~1A and~1B and 80~GB for Case~3.

\subsection{Case 1A: observations through year~6}
\label{sec:3t-sample9}

\begin{figure}[!htbp]
\centering
\includegraphics[width=\linewidth]
{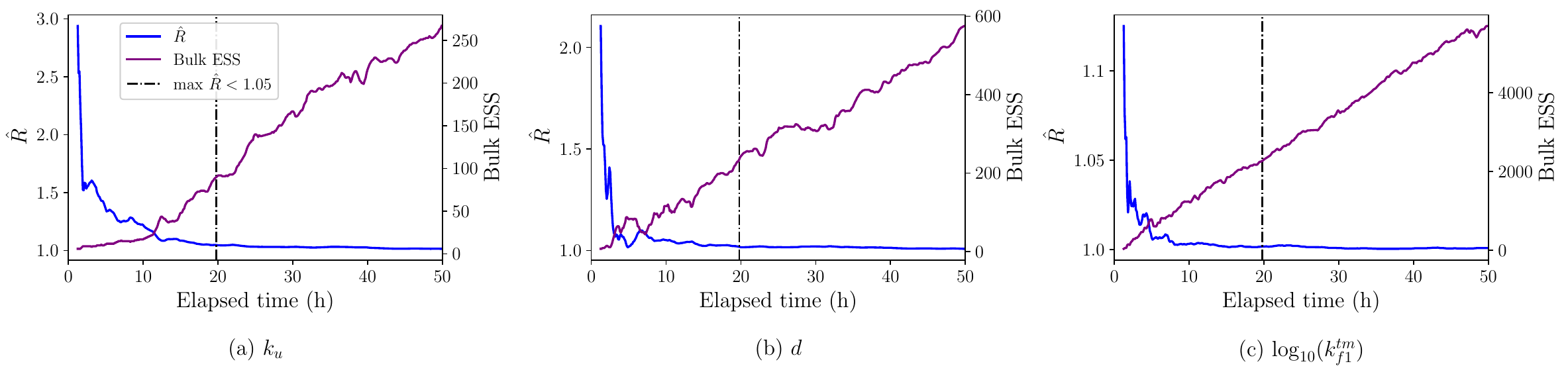}

\caption{MCMC convergence diagnostics for Case~1A. The $\widehat{R}$ statistic (left axis) and bulk effective sample size (right axis) are shown as functions of elapsed time for three metaparameters: (a) $k_u$ (minimum bulk ESS), (b) coefficient $d$ (median bulk ESS), and (c) $\log_{10}(k_{f1}^{\mathrm{tm}})$ (maximum bulk ESS), based on the values at the end of the 50-hour run. The vertical dashed line at 19.8~hours indicates the satisfaction of the convergence criterion $\max_{j=1,\ldots,11}\widehat{R}_j<1.05$. The MCMC calculation terminates at 50~hours upon reaching the time limit specified in the stopping rule.}
\label{fig:3t-s9-rhat-case1}
\end{figure}

Because MCMC posterior estimates provide a reference for evaluating SMC, we first assess MCMC convergence following the procedure described in Section~\ref{sec:mcmc}. Fig.~\ref{fig:3t-s9-rhat-case1} shows the evolution of the $\widehat{R}$ statistic and bulk effective sample size (bulk ESS) for three representative metaparameters. The bulk ESS values are computed using the rank-normalized split-chain procedure described in Section~\ref{sec:mcmc}, applied to the five MCMC chains after burn-in removal. The convergence criterion $\max_{j=1,\ldots,11}\widehat{R}_j<1.05$ is satisfied after 19.8~hours of computation. The MCMC chains are then continued until the 50-hour time limit is reached (this time limit is reached before the $\max_{j}\widehat{R}_j<1.01$ stopping rule is achieved). Based on the bulk ESS values at the end of the run, $k_u$ (permeability of the upper aquifer), $d$ (coefficient relating porosity and log-permeability), and $\log_{10}(k_{f1}^{tm})$ (permeability of the lower portion of fault~1) correspond to the minimum, median, and maximum bulk ESS values, respectively, among the 11 metaparameters. These three quantities thus illustrate the range of MCMC sampling efficiency. For all three metaparameters, $\widehat{R}$ decreases quickly during the early stage of sampling and gradually approaches~1. The convergence of $k_u$ is slower, with $\widehat{R}$ continuing to decrease over about the first 20~hours and bulk ESS reaching $\sim$250 at the end of the run. The most efficient sampling is observed for $\log_{10}(k_{f1}^{tm})$. In this case, $\widehat{R}$ approaches~1 within the first several hours, and bulk ESS increases nearly monotonically to $\sim$5000. The results for $k_u$, $d$, and $\log_{10}(k_{f1}^{tm})$ demonstrate the large variation in sampling efficiency over the metaparameters.

The SMC calculation for Case~1A uses $M=30$ blockwise MCMC mutation steps per particle at each intermediate tempering iteration. Throughout the calculation, we monitor the tempering exponent $\beta_k$, the importance-weight ESS, and the mean acceptance rates for the latent variable and metaparameter updates. The SMC calculation terminates after 13 tempering iterations when $\beta_K=1$, at which point the full likelihood has been incorporated and the resulting weighted particles approximate the posterior distribution. Unlike MCMC, SMC does not require a separate burn-in period or an $\widehat{R}$-based convergence criterion.

\begin{figure}[!htbp]
\centering
\includegraphics[width=\linewidth]{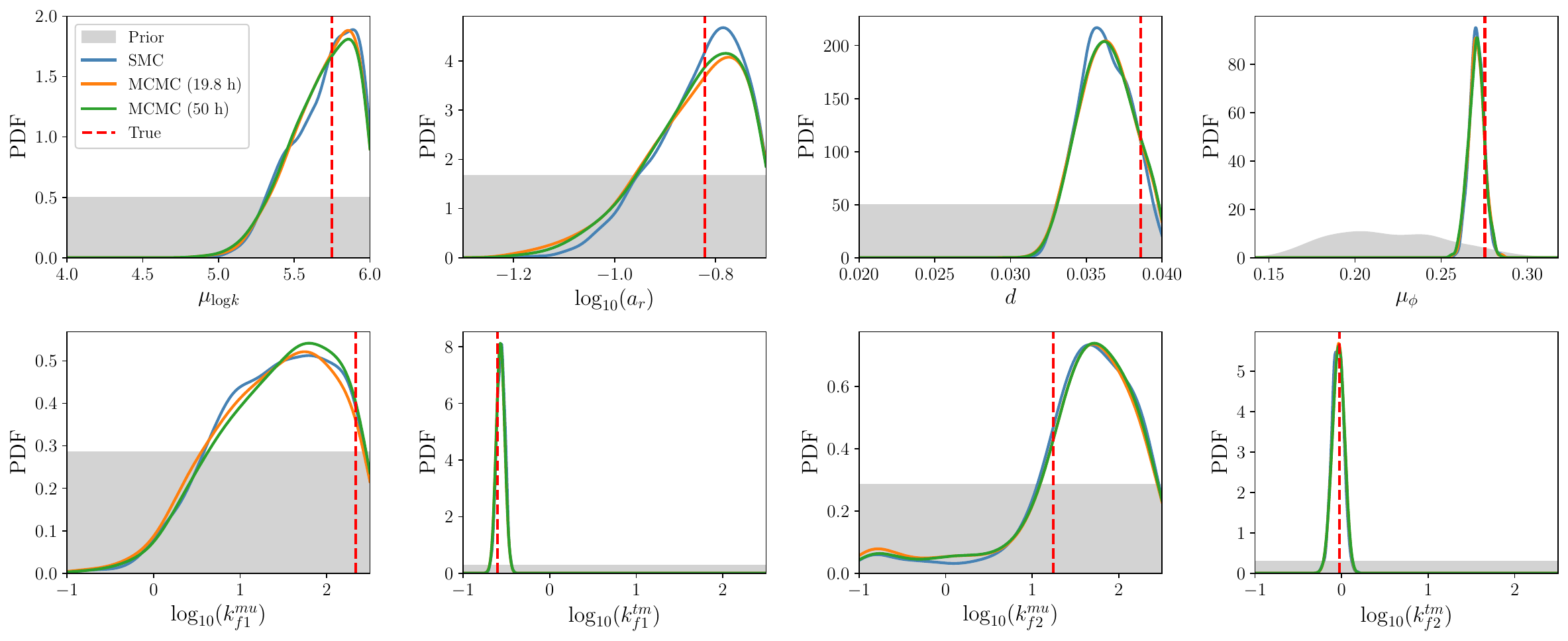}
\caption{Marginal posterior distributions for Case~1A. Results are shown for seven metaparameters with MCMC bulk ESS values above 450 at the end of the 50-hour run, along with the derived mean porosity $\mu_{\phi}$. Distributions are shown for SMC (blue), MCMC at the first satisfaction of $\max_{j=1,\ldots,11}\widehat{R}_j<1.05$ (19.8~hours, orange), and MCMC at the time-limit stopping point (50~hours, green). Prior distributions are shown in gray, and the true values are indicated by red dashed vertical lines.}
\label{fig:3t-s9-marg}
\end{figure}

The marginal posterior distributions of key metaparameters for Case~1A, obtained using SMC and MCMC, are compared in Fig.~\ref{fig:3t-s9-marg}. MCMC results at both 19.8~hours (when the convergence criterion is first satisfied) and 50~hours (when the time-limit condition of the stopping rule is met) are shown. The marginal densities in this figure are estimated using kernel density estimation (KDE), with the bandwidth determined by Scott's rule ($h=h_{\mathrm{Scott}}$). The metaparameters above a bulk ESS value of 450 generally exhibit greater prior-to-posterior uncertainty reduction than those below this value. We therefore denote these two groups of metaparameters as (relatively) well constrained versus weakly constrained by the observations, respectively. Fig.~\ref{fig:3t-s9-marg} includes results for the seven well-constrained metaparameters, for which the MCMC bulk ESS values exceed 450 at the end of the 50-hour run (we note that an average ESS of at least 50 per split chain was recommended in \citep{Vehtari2021}, which would correspond to a combined ESS of 500 for the five chains used here, though we apply the slightly lower value of 450 as a descriptive cutoff). The derived mean porosity $\mu_{\phi}$ is also presented in Fig.~\ref{fig:3t-s9-marg}.

The close agreement between the two sets of MCMC results indicates that the marginal distributions remain stable under additional sampling. Importantly, the SMC distributions are seen to closely coincide with the MCMC results. There are small discrepancies in some cases (e.g., for $\log_{10}a_r$), though the overall correspondence between the SMC and reference MCMC results is excellent. It is also worth noting that the true values lie within the posterior distributions in all cases. Although uncertainty reduction is clearly observed for all eight quantities, the posterior distributions for $\log_{10}(k_{f1}^{mu})$ and $\log_{10}(k_{f2}^{mu})$ remain relatively broad.

\begin{figure}[!htbp]
\centering
\includegraphics[width=\linewidth]{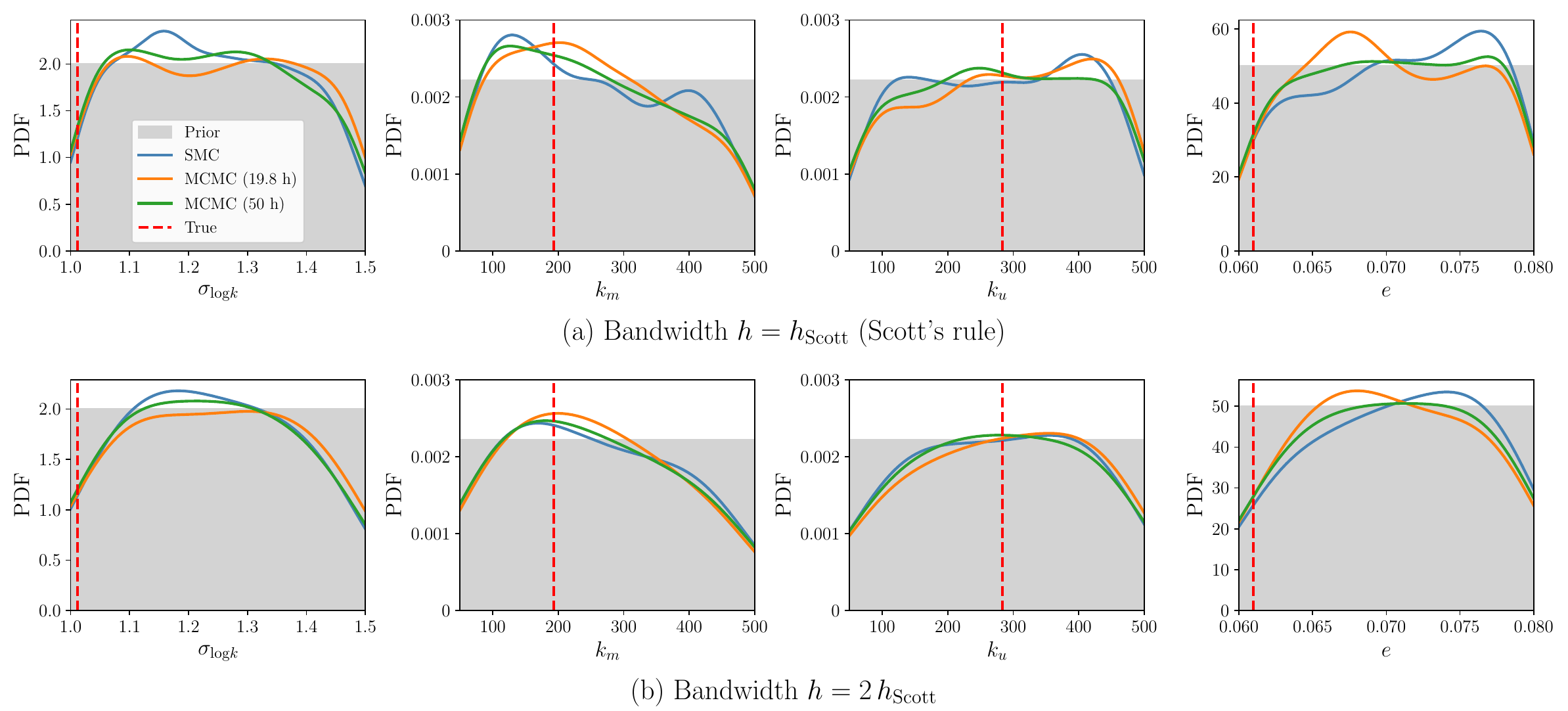}
\caption{Bandwidth sensitivity of kernel density estimates for the marginal posterior distributions of $\sigma_{\log k}$, $k_m$, $k_u$, and $e$ in Case~1A. These four weakly constrained metaparameters have MCMC bulk ESS values below 450 at the end of the 50-hour run. The upper row uses Scott's rule, $h=h_{\mathrm{Scott}}$, and the lower row uses $h=2h_{\mathrm{Scott}}$. Results are shown for SMC (blue), MCMC at the satisfaction of $\max_{j=1,\ldots,11}\widehat{R}_j<1.05$ (19.8~hours, orange), and MCMC at the time-limit stopping point (50~hours, green). Gray region denotes the prior distributions, and the red dashed vertical lines indicate the true parameter values.}
\label{fig:3t-s9-marg-bw}
\end{figure}

The remaining four metaparameters, $\sigma_{\log k}$, $k_m$, $k_u$, and $e$, have MCMC bulk ESS values below 450 at the end of the 50-hour run and show little uncertainty reduction relative to their priors. These parameters are weakly constrained by the observations, and the relatively low bulk ESS values suggest that finite-sample fluctuations may affect the estimated density shapes, particularly in terms of small local peaks. To examine the sensitivity of these features to bandwidth selection, in Fig.~\ref{fig:3t-s9-marg-bw} we compare KDEs obtained using Scott's rule ($h=h_{\mathrm{Scott}}$) to those using twice the Scott bandwidth ($h=2h_{\mathrm{Scott}}$). The local peaks observed with Scott's rule are eliminated with the larger bandwidth, while the overall shapes of the broad posterior distributions are maintained. This suggests that the minor local peaks should be interpreted cautiously rather than regarded as well-resolved posterior modes. For consistent presentation and comparison, we use $h=2h_{\mathrm{Scott}}$ for the SMC and MCMC marginal density estimates in all subsequent cases. This has very little effect on the level of agreement between the two methods for the well-constrained metaparameters shown in Fig.~\ref{fig:3t-s9-marg}.

\begin{figure}[!htbp]
\centering
\includegraphics[width=\linewidth]{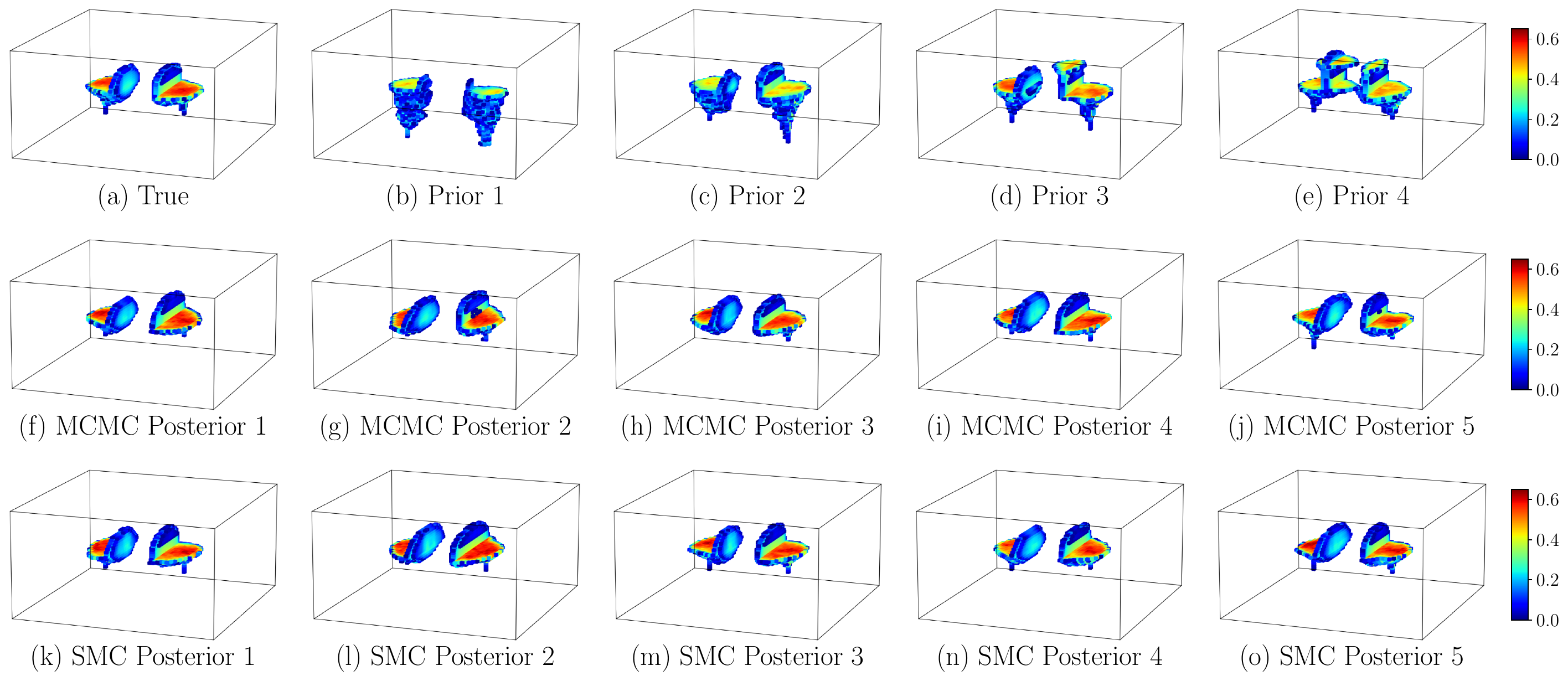}
\caption{Representative 3D CO$_2$ saturation plumes at 50~years for Case~1A. Top row shows the true model and four representative prior realizations, and middle and bottom rows show representative MCMC and SMC posterior samples. Compared to the prior plumes, both sets of posterior results show less variability and general agreement with the true plume geometry.}
\label{fig:3t-s9-plumes}
\end{figure}

We next consider posterior predictions of 3D CO$_2$ saturation distributions at 50~years for Case~1A. Fig.~\ref{fig:3t-s9-plumes} shows the true CO$_2$ plume, along with representative prior and posterior realizations selected by clustering the corresponding saturation fields using the procedure described in \citep{jiang2024History}. Here and in subsequent figures for Case~1A, the MCMC results are from predictions after 19.8~hours of computation (which are very close to those at 50~hours). The prior realizations (Fig.~\ref{fig:3t-s9-plumes}b--e) show large variation in plume geometry and in the extent of CO$_2$ migration through the faults. The representative MCMC and SMC posterior samples (Fig.~\ref{fig:3t-s9-plumes}f--o), by contrast, exhibit much less variation and are in general correspondence with the true plume. The posterior saturation fields capture the limited amount of vertical CO$_2$ migration. These results indicate that early-time observations provide useful information for long-term plume migration predictions in Case~1A, and that SMC accurately captures the posterior uncertainty in the overall CO$_2$ plume geometry.

\begin{figure}[!htbp]
\centering
\includegraphics[width=0.95\linewidth]{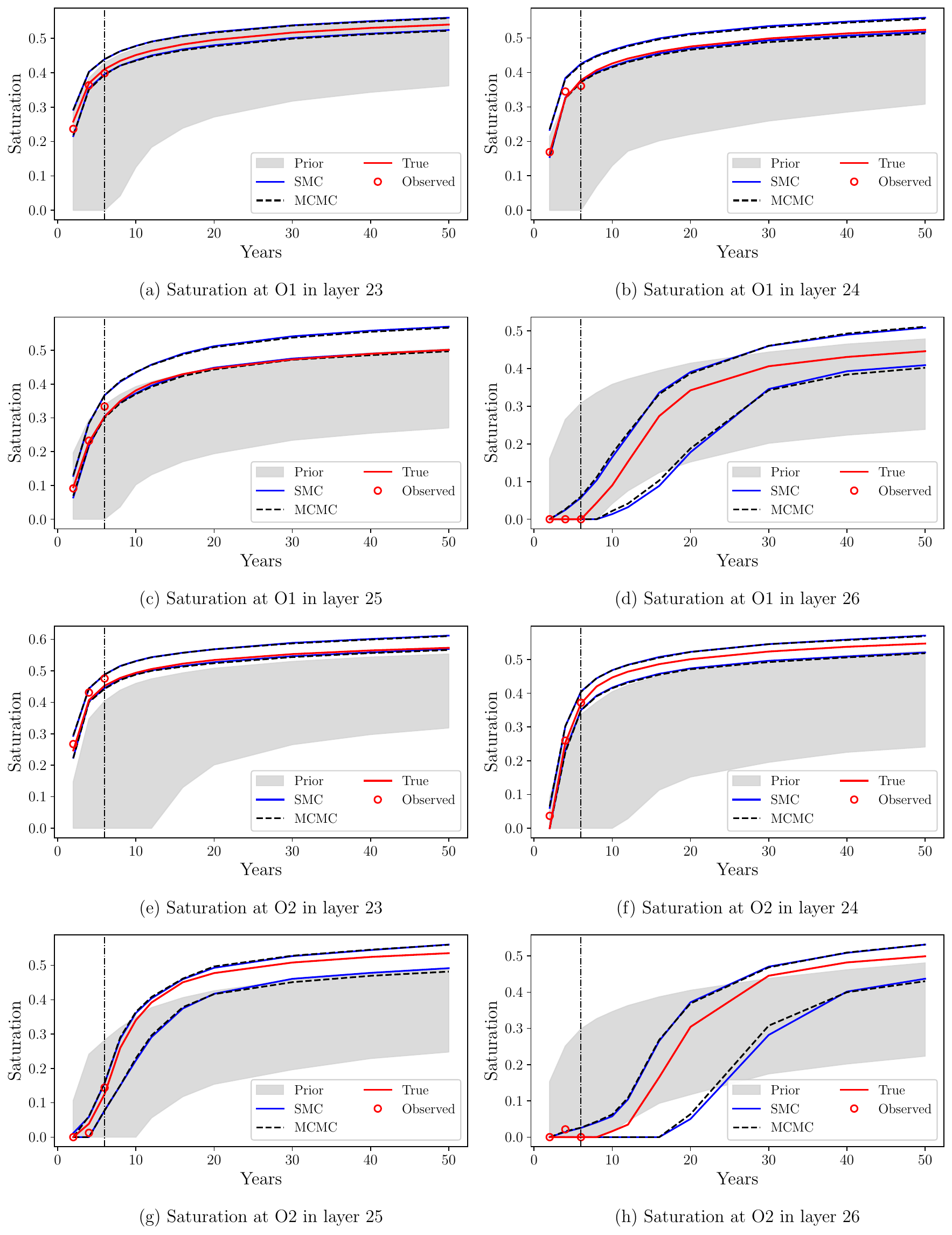}
\caption{Posterior $\mathrm{P}_{10}$--$\mathrm{P}_{90}$ predictions of CO$_2$ saturation at observation wells O1 and O2 in various layers for Case~1A. Results are shown for SMC (blue) and MCMC (black dashed), along with the prior $\mathrm{P}_{10}$--$\mathrm{P}_{90}$ range (gray), true response (red), and observed data (red circles). Vertical dashed line indicates the end of the data assimilation time window at 6~years.}
\label{fig:3t-s9-wells-sw}
\end{figure}

Next, the detailed saturation responses at the observation wells are considered. Fig.~\ref{fig:3t-s9-wells-sw} shows the prior $\mathrm{P}_{10}$--$\mathrm{P}_{90}$ range (10th to 90th percentile, shown in gray), along with the  $\mathrm{P}_{10}$--$\mathrm{P}_{90}$ posterior predictions for CO$_2$ saturation at observation wells~O1 and O2 in particular layers. The prior ranges are large, and the true responses (red curves) are at the high end of the prior at many locations. The observed data are indicated by the red circles, which deviate from the red curves due to the added noise. Large uncertainty reduction is achieved in all locations except for those in layer~26. Importantly, the $\mathrm{P}_{10}$--$\mathrm{P}_{90}$ SMC and MCMC posterior predictions are in close agreement over the full 50-year period at all locations. The larger posterior uncertainty is observed at locations in layer~26 that have not experienced CO$_2$ arrival over the 6-year data assimilation time window (Fig.~\ref{fig:3t-s9-wells-sw}d and h), as would be expected.

\begin{figure}[!htbp]
\centering
\includegraphics[width=0.9\linewidth]{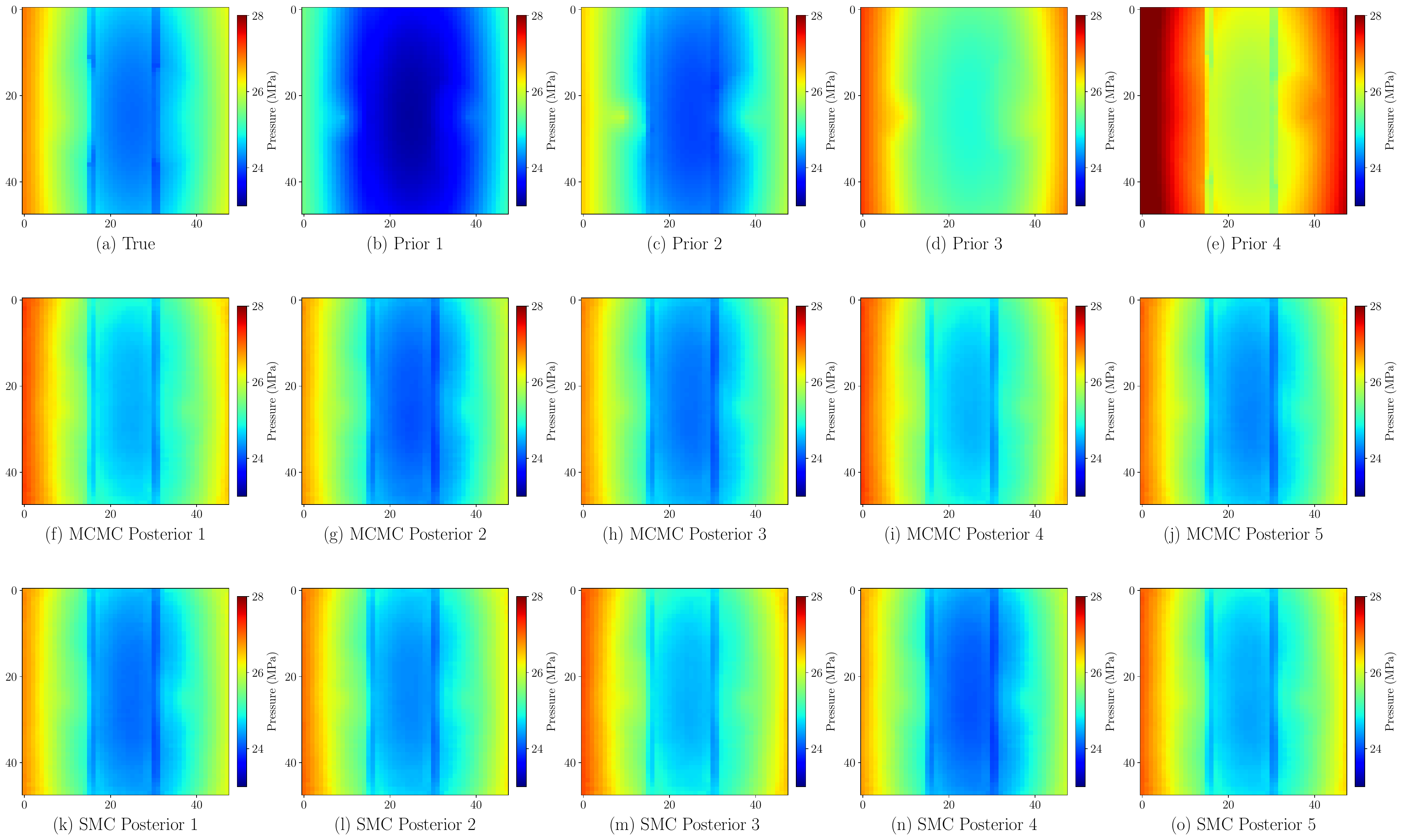}
\caption{Pressure maps in the top layer of the storage aquifer at 50~years for Case~1A. Top row shows the true model and four representative prior realizations, and the middle and bottom rows show representative MCMC and SMC posterior samples.}
\label{fig:3t-s9-pmap}
\end{figure}

Results for pressure are now presented. We display 2D maps rather than 3D fields for pressure since key details are easier to see in the 2D images. Fig.~\ref{fig:3t-s9-pmap} shows representative (clustered) pressure maps in the top layer of the storage aquifer at 50~years. The true model response appears in Fig.~\ref{fig:3t-s9-pmap}a. Higher pressures are observed at the left and right edges due to aquifer dip. The impact of the faults is also visible in Fig.~\ref{fig:3t-s9-pmap}a. The prior realizations (Fig.~\ref{fig:3t-s9-pmap}b--e) show substantial variability in pressure magnitude, spatial variation, and the impact of the faults. Data assimilation of observations through year~6 greatly reduces this variability. The representative MCMC and SMC posterior maps capture the key features and the general pressure range of the true field. Some variation remains in the magnitude and extent of the central lower-pressure region, which reflects posterior uncertainty. General agreement between the MCMC and SMC results is again observed.

\begin{figure}[!htbp]
\centering
\includegraphics[width=0.9\linewidth]{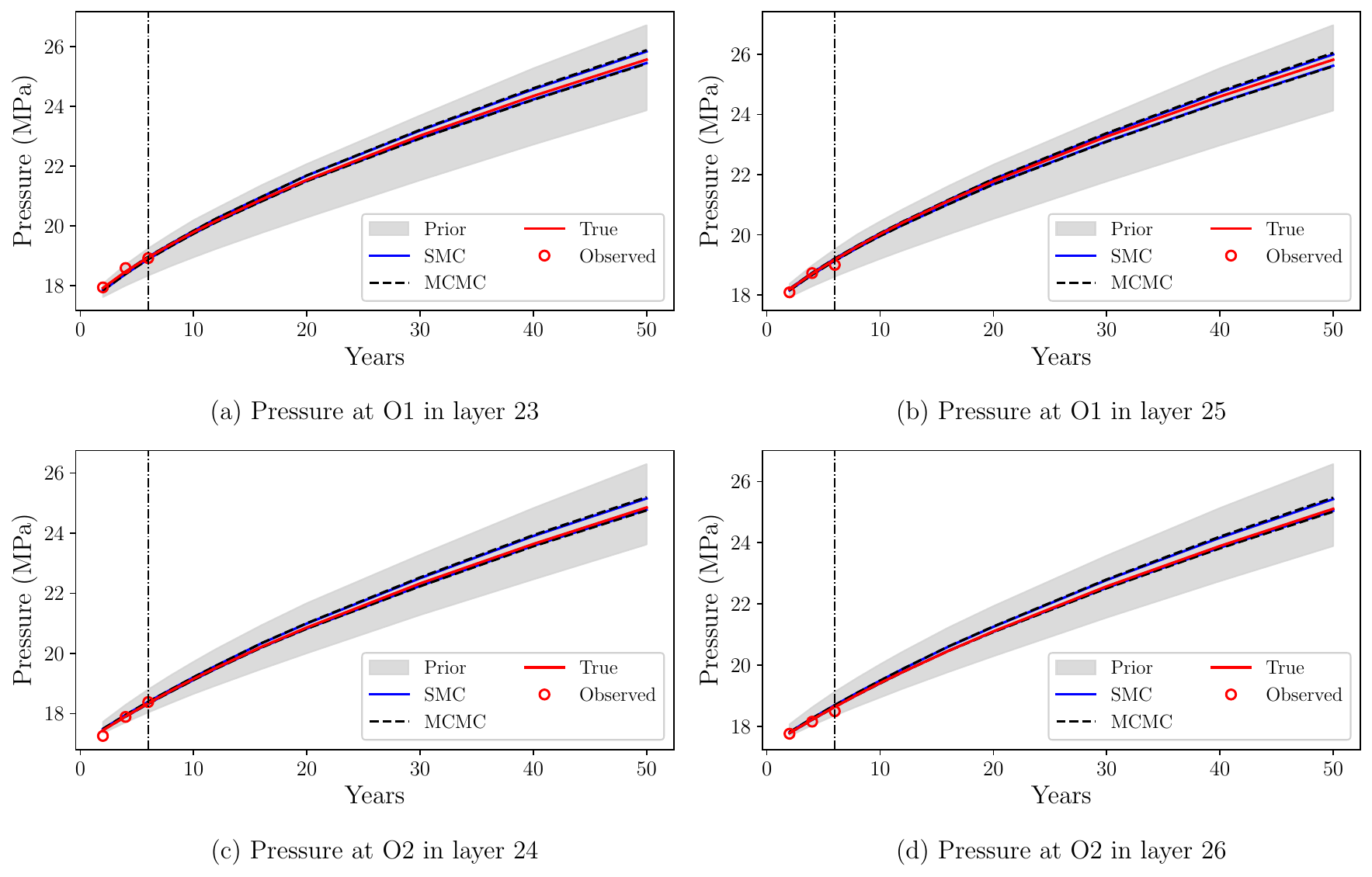}
\caption{Posterior $\mathrm{P}_{10}$--$\mathrm{P}_{90}$ predictions of pressure at observation wells O1 and O2 in various layers for Case~1A. Results are shown for SMC (blue) and MCMC (black dashed), along with the prior $\mathrm{P}_{10}$--$\mathrm{P}_{90}$ range (gray), true response (red), and observed data (red circles). Vertical dashed line indicates the end of the data assimilation time window at 6~years.}
\label{fig:3t-s9-wells-p}
\end{figure}

Posterior $\mathrm{P}_{10}$--$\mathrm{P}_{90}$ predictions for pressure in various layers at both observation wells are shown in Fig.~\ref{fig:3t-s9-wells-p}. Substantial uncertainty reduction is achieved in all layers at both wells, and the posterior intervals capture the observed data. The SMC and MCMC predictions are again very close.

The computational costs of MCMC and SMC for the faulted carbon storage system are summarized in Table~\ref{tab:seam-computational-efficiency}. Here we include results for a second true model, denoted as Case~3 (detailed posterior metaparameter and flow results for Case~3 are provided in SM). The timings given in Table~\ref{tab:seam-computational-efficiency} correspond to MCMC-to-SMC elapsed-time ratios of approximately 3.0 and 2.6. We also see that SMC requires an average of $\sim$34\% fewer function evaluations than MCMC. The MCMC timings and number of function evaluations in Table~\ref{tab:seam-computational-efficiency} are for the less stringent convergence criterion ($\max_{j=1,\ldots,11}\widehat{R}_j<1.05$). For both Cases~1A and~3, the MCMC calculations were continued to the 50-hour time limit without satisfying the more stringent criterion ($\max_{j=1,\ldots,11}\widehat{R}_j<1.01$). Based on 50-hour MCMC runtimes, the speedups using SMC would correspond to 7.6 and 6.4.

\begin{table}[!htbp]
\centering
\small
\caption{Computational cost comparison for MCMC and SMC for the faulted carbon storage system (Cases~1A and~3). Elapsed times and numbers of function evaluations are reported for single-GPU calculations.}
\label{tab:seam-computational-efficiency}
\begin{tabular*}{0.95\linewidth}{@{\extracolsep{\fill}}lrrrr@{}}
\toprule
& \multicolumn{2}{c}{Case~1A}
& \multicolumn{2}{c}{Case~3} \\
\cmidrule(lr){2-3}\cmidrule(lr){4-5}
Quantity & MCMC & SMC & MCMC & SMC \\
\midrule
Elapsed time (h) & 19.8 & 6.6 & 20.3 & 7.8 \\
Function evaluations
& $804{,}960$ & $515{,}523$
& $1{,}001{,}230$ & $687{,}590$ \\
\bottomrule
\end{tabular*}
\end{table}

In our MCMC implementation, surrogate model evaluations for the five independent chains are performed in batches on a single GPU, thus exploiting parallelism across chains. Further acceleration is limited by the sequential updates within each chain and the relatively small number of chains (i.e., the use of multiple GPUs has no impact on MCMC elapsed time for Cases~1A and~3). SMC, by contrast, involves a much larger population of particles, for which surrogate evaluations and MCMC mutation updates can be performed independently across particles. This enables SMC to make effective use of multiple GPUs. Along these lines, we performed parallel SMC calculations for Cases~1A and~3 using five GPUs in each case. The same setups as in the corresponding single-GPU calculations were otherwise specified. The parallel SMC runs required 1.4 and 1.7~hours, respectively, corresponding to speedups of about 4.7 and 4.6 relative to single-GPU SMC, and 14.1 and 11.9 relative to single-GPU MCMC (Table~\ref{tab:seam-computational-efficiency}). This demonstrates the substantial speed advantage relative to MCMC that SMC can provide when multiple GPUs are available.

We also assessed sampling efficiency in terms of the Monte Carlo precision of posterior mean estimates per unit elapsed time. Following~\citet{HerbstSchorfheide2014} and \citet{FanLeslieWand2008}, sampling efficiency for a metaparameter $h_j$ is defined as $\mathcal{E}_{h_j} = \operatorname{Var}_{\pi}(h_j)/[\operatorname{Var}_{\mathrm{MC}}(\widehat{\mu}_{h_j})\,T]$, where $\operatorname{Var}_{\mathrm{MC}}(\widehat{\mu}_{h_j})$ is the variance of the estimated posterior mean across independent runs and $T$ is the elapsed time. This measure provides a consistent basis for comparing SMC and MCMC. Comparisons based on five independent runs of each method suggest generally higher sampling efficiency for SMC, consistent with these previous studies.

Taken in total, the results for Case~1A demonstrate that SMC provides posterior estimates in close agreement with MCMC. The computational comparisons for Cases~1A and~3 demonstrate clear cost reduction with SMC, which is quite substantial when multi-GPU parallelization is applied. We next consider another advantage of SMC, namely, its ability to perform successive data assimilation as additional observations become available.

\subsection{Case~1B: observations through year~12}
\label{sec:6t-sample9}

Case~1B considers the assimilation of additional observations at 8, 10, and 12~years. Successive SMC updates the existing posterior particles obtained at year~6, whereas the MCMC procedure used here requires a new calculation initialized from the joint prior using all observations through year~12. A single-stage SMC calculation initialized from the prior that uses all observations is also performed.

The single-stage SMC calculation requires 17 tempering iterations, with $M=50$ blockwise MCMC mutation steps per particle at each intermediate iteration. The first stage of successive SMC corresponds to the Case~1A calculation described in Section~\ref{sec:3t-sample9}, which requires 13 tempering iterations with $M=30$. Starting from the posterior particles at year 6, the successive SMC update assimilates additional observations at 8, 10, and 12~years in six tempering iterations with $M=40$. The MCMC, single-stage SMC, and successive SMC calculations each use a single NVIDIA A100 GPU with 40~GB of memory.

\begin{figure}[!tp]
\centering
\includegraphics[width=0.92\linewidth]{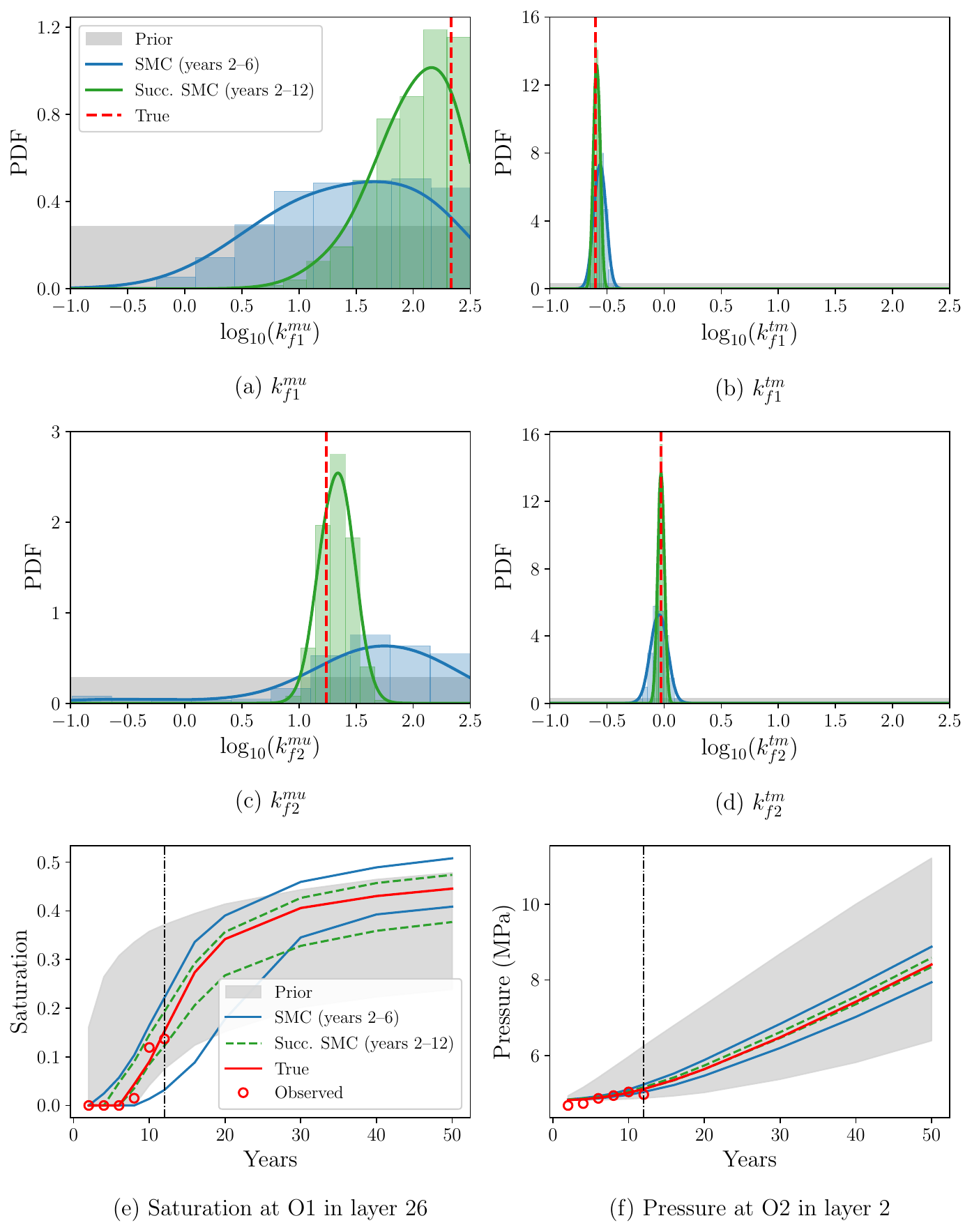}
\caption{SMC posterior based on observations through year~6 (blue) compared with successive SMC posterior after assimilation of data through year~12 (green). (a)--(d): Results for fault metaparameters, with uniform prior in gray and true values indicated by vertical lines. Additional observations lead to uncertainty reduction in all four metaparameters. (e) and (f): Posterior $\mathrm{P}_{10}$--$\mathrm{P}_{90}$ predictions of CO$_2$ saturation and pressure based on observations through year~6 (blue) and after a successive SMC update through year~12 (dashed green). Prior $\mathrm{P}_{10}$--$\mathrm{P}_{90}$ range (gray regions), true response (red curves), and observations (red circles) are also shown. Vertical dashed line indicates the end of the assimilation window at 12~years.}
\label{fig:s9-seq-faults}
\end{figure}

Here we focus on the additional uncertainty reduction achieved using more data and on predictions of CO$_2$ saturation and pressure at observation well locations. Results for the fault-permeability metaparameters (plotted using $h=2h_{\mathrm{Scott}}$, which differs from the value used in Fig.~\ref{fig:3t-s9-marg}) from successive assimilation of observations at 8, 10, and 12~years are shown in Fig.~\ref{fig:s9-seq-faults}a--d. Relative to the posterior distributions obtained at year~6, we see that the updated distributions for all fault quantities are more concentrated near their true values. These metaparameters are sensitive to pressure communication and CO$_2$ migration between the target, middle, and upper aquifers, and the additional 6~years of data appear to be quite informative. Uncertainty reduction is particularly clear for $\log_{10}(k_{f1}^{mu})$ and $\log_{10}(k_{f2}^{mu})$, which are broadly distributed after assimilation at year~6, but become more concentrated after the successive update. The posterior predictive ranges, shown in Fig.~\ref{fig:s9-seq-faults}e and f, are also reduced. Specifically, the successive update narrows the posterior $\mathrm{P}_{10}$--$\mathrm{P}_{90}$ prediction for saturation in layer~26 at O1 and for pressure in layer~2 at O2. Both sets of results are in close alignment with the true response.

\begin{figure}[!tp]
\centering
\includegraphics[width=0.93\linewidth]{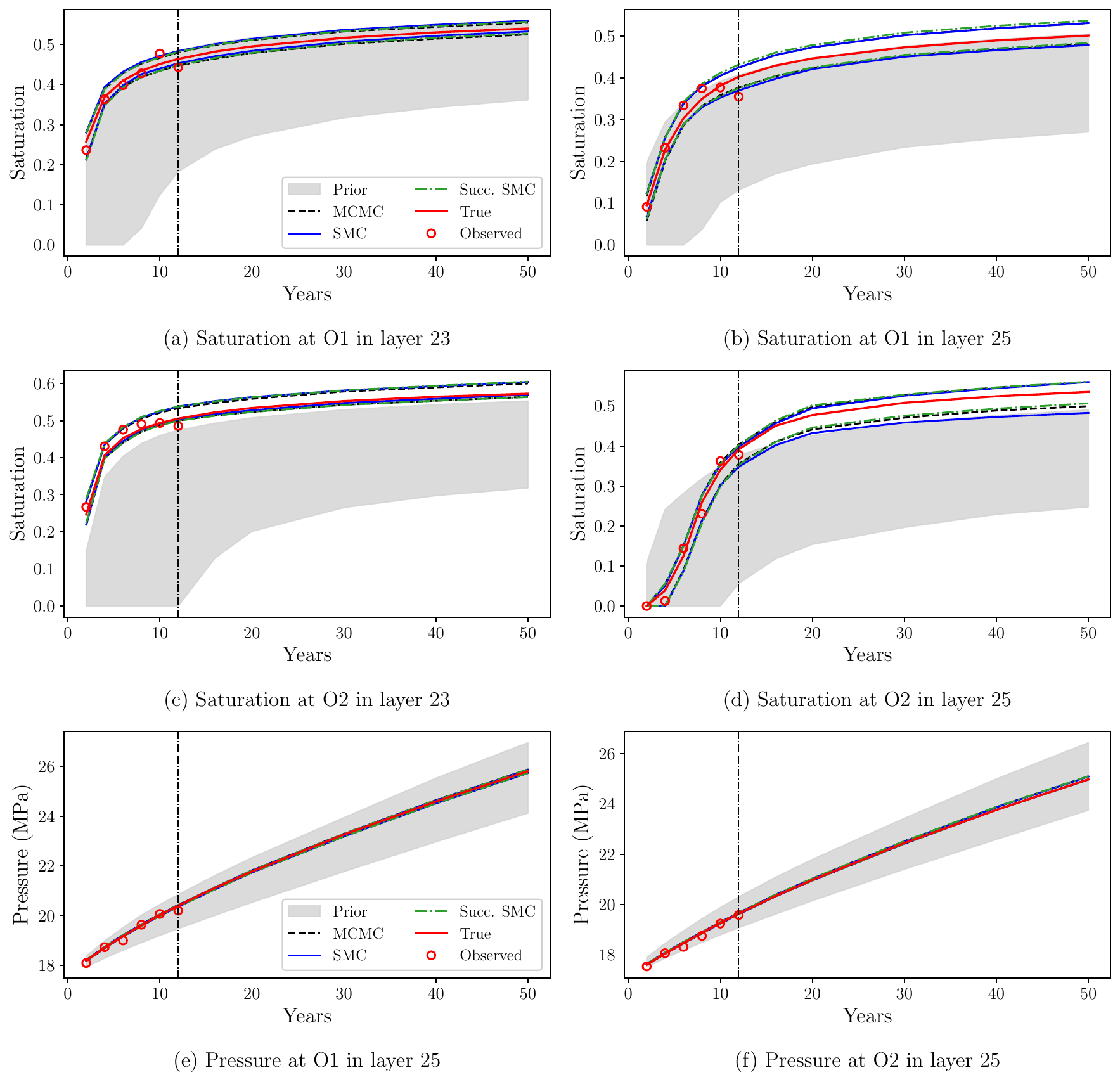}
\caption{Posterior $\mathrm{P}_{10}$--$\mathrm{P}_{90}$ predictions of CO$_2$ saturation and pressure in various layers at observation wells for Case~1B. Results are shown for single-stage SMC (blue), MCMC from the 50-hour run (black dashed), and successive SMC (green dashed), along with the prior $\mathrm{P}_{10}$--$\mathrm{P}_{90}$ range (gray), true response (red), and observed data (red circles). Vertical dashed line indicates the end of the data assimilation time window at year~12.}
\label{fig:6t-s9-wells-sw}
\end{figure}

Results using successive SMC, single-stage SMC, and MCMC are compared, in terms of posterior $\mathrm{P}_{10}$--$\mathrm{P}_{90}$ predictions of CO$_2$ saturation and pressure at observation well locations, in Fig.~\ref{fig:6t-s9-wells-sw}. The single-stage SMC and MCMC calculations are initialized from the prior and use all observations through year~12. All three approaches substantially reduce the prior predictive uncertainty, and the posterior predictions are generally consistent with the observed data during the assimilation period. The single-stage SMC, MCMC, and successive SMC predictions are in close agreement over the full 50-year simulation period, and the true responses generally lie within the posterior $\mathrm{P}_{10}$--$\mathrm{P}_{90}$ ranges. The marginal posterior distributions for the metaparameters obtained using the three approaches are compared in Fig.~S2 in SM. The three sets of results are in reasonable agreement, with the more noticeable differences generally occurring for metaparameters that display limited amounts of uncertainty reduction (e.g., $k_u$ and $e$).

The computational costs for Case~1B are summarized in Table~\ref{tab:case1b-cost}. The MCMC calculation terminates at the 50-hour time limit specified in the stopping rule in Section~\ref{sec:mcmc}, without satisfying the convergence criterion $\max_{j=1,\ldots,11}\widehat{R}_j<1.05$. By contrast, the total cost of both stages of the successive SMC procedure is only 11.5~hours. These results further demonstrate the potential advantages of our SMC procedure for problems of this type.

\begin{table}[!htbp]
\centering
\small
\caption{Computational cost comparison for Case~1B. For successive SMC, both the incremental update cost and the total cost including the initial Case~1A calculation are reported.}
\label{tab:case1b-cost}

\setlength{\tabcolsep}{4pt}
\renewcommand{\arraystretch}{1.1}
\renewcommand{\tabularxcolumn}[1]{m{#1}}

\begin{tabularx}{\linewidth}{@{}l@{\hspace{22pt}} >{\raggedright\arraybackslash}X rr@{}}
\toprule
Approach & Description & Elapsed time (h) & Function evaluations \\
\midrule
MCMC & Rerun from the start using all observations
     & 50.0 & $1{,}807{,}005$ \\
\addlinespace[3pt]
SMC & Rerun from the start using all observations
    & 17.4 & $1{,}246{,}943$ \\
\addlinespace[5pt]
Successive SMC & Update year-6 posterior using new observations
               & 4.9 & $382{,}426$ \\
\addlinespace[3pt]
Successive SMC & Total cost for Case~1A and the successive update
               & 11.5 & $897{,}949$ \\
\bottomrule
\end{tabularx}
\end{table}

Additional results for Case~3, which considers a different synthetic true model with distinct CO$_2$ plume geometry and pressure distribution, are provided in Figs.~S3--S8 in SM. For this case, SMC and MCMC again provide closely matching posterior metaparameter distributions and flow predictions, with substantial reduction in predictive uncertainty relative to the prior. Taken in total, the results for Cases~1A, 1B, and~3 demonstrate that SMC provides accurate posterior estimates at reduced computational cost, while successive SMC enables efficient assimilation of additional observations by updating existing posterior particles.

\section{Data assimilation results for the channelized oil-water system}
\label{sec:oilwater_results}

In this section, we apply our data assimilation procedures to the channelized oil-water system described in Section~\ref{sec:oil_geomodel}. The uncertain quantities include the spatial facies distribution represented using the 3D-LDM parameterization and the facies-dependent petrophysical properties. Results are evaluated in terms of SMC and MCMC posterior petrophysical and geometric metaparameter distributions, representative facies realizations, well and field rate predictions, and computational efficiency. We consider two synthetic true models, referred to as Case~2 and Case~4. Detailed results for Case~2 are presented in this section. Most of the results for Case~4 are provided in SM, though timing information is presented in tabular form in this section.

For Cases~2 and~4, the observed data comprise oil and water production rates at the $n_{\mathrm{prod}}=6$ producers (P1--P6) and water injection rates at the $n_{\mathrm{inj}}=3$ injectors (I1--I3). Observations are collected at 50-day intervals through day 250, giving five observation times and a total of $n_m=5(2n_{\mathrm{prod}}+n_{\mathrm{inj}})=75$ measurements. We consider synthetic observed data generated using simulated rates for the true model (shown later) combined with Gaussian measurement noise. The standard deviation of the measurement error is set to 5\% of the corresponding true rate. Posterior predictions are evaluated over the full 1500-day simulation period.

The MCMC calculations follow the procedure described in Section~\ref{sec:mcmc}, with five independent chains and the first 5000 iterations of each chain discarded as burn-in. For Cases~2 and~4, SMC uses $N=1000$ particles, $M=30$ blockwise MCMC mutation steps per particle at each intermediate tempering iteration, and a fixed importance-weight ESS fraction of $\rho=0.5$. Each MCMC and SMC calculation uses a single NVIDIA A100 GPU with 40~GB of memory. The SMC calculations terminate after nine and ten tempering iterations for Cases~2 and~4, respectively, when $\beta_K=1$.

\begin{figure}[!htbp]
\centering
\includegraphics[width=\linewidth]
{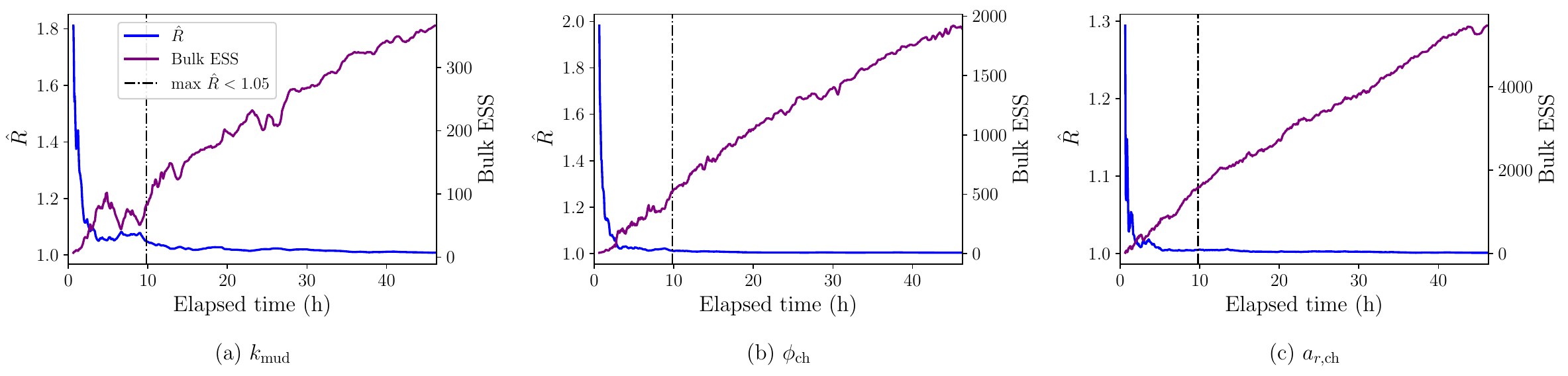}
\caption{MCMC convergence diagnostics for Case~2. The $\widehat{R}$ statistic (left axis) and bulk effective sample size (bulk ESS; right axis) are shown as functions of elapsed time for three metaparameters. (a)~$k_{\mathrm{mud}}$, which has the minimum ESS, (b)~$\phi_{\mathrm{ch}}$, which has the median ESS, and (c)~$a_{r,\mathrm{ch}}$, which has the maximum ESS. Vertical dashed line at 9.8~hours indicates satisfaction of the convergence criterion $\max_{j=1,\ldots,9}\widehat{R}_j<1.05$. The MCMC run terminates at 46.3~hours when the more stringent criterion, $\max_{j=1,\ldots,9}\widehat{R}_j<1.01$, is satisfied.}
\label{fig:tm411-rhat-ess}
\end{figure}

MCMC results again provide a reference for SMC. We first assess MCMC convergence following the procedure described in Section~\ref{sec:mcmc}. Fig.~\ref{fig:tm411-rhat-ess} shows the evolution of the $\widehat{R}$ statistic and bulk effective sample size (bulk ESS) for the petrophysical metaparameters corresponding to the minimum ($k_{\mathrm{mud}}$), median ($\phi_{\mathrm{ch}}$), and maximum ($a_{r,\mathrm{ch}}$) bulk ESS values. The convergence criterion $\max_j\widehat{R}_j<1.05$ is satisfied after 9.8~hours. The MCMC chains are then continued until the more stringent criterion $\max_j\widehat{R}_j<1.01$ is satisfied, which occurs at 46.3~hours. For all three metaparameters, $\widehat{R}$ decreases rapidly during the early stage of sampling and gradually approaches 1.0, while bulk ESS increases with elapsed time. For the parameter with the slowest sampling ($k_{\mathrm{mud}}$), bulk ESS is $\sim$300 at the end of the run. By contrast, $a_{r,\mathrm{ch}}$ exhibits much more efficient sampling, with bulk ESS reaching $\sim$5000. These results indicate substantial variation in sampling efficiency among the petrophysical metaparameters.

\begin{figure}[!htbp]
\centering
\includegraphics[width=\linewidth]{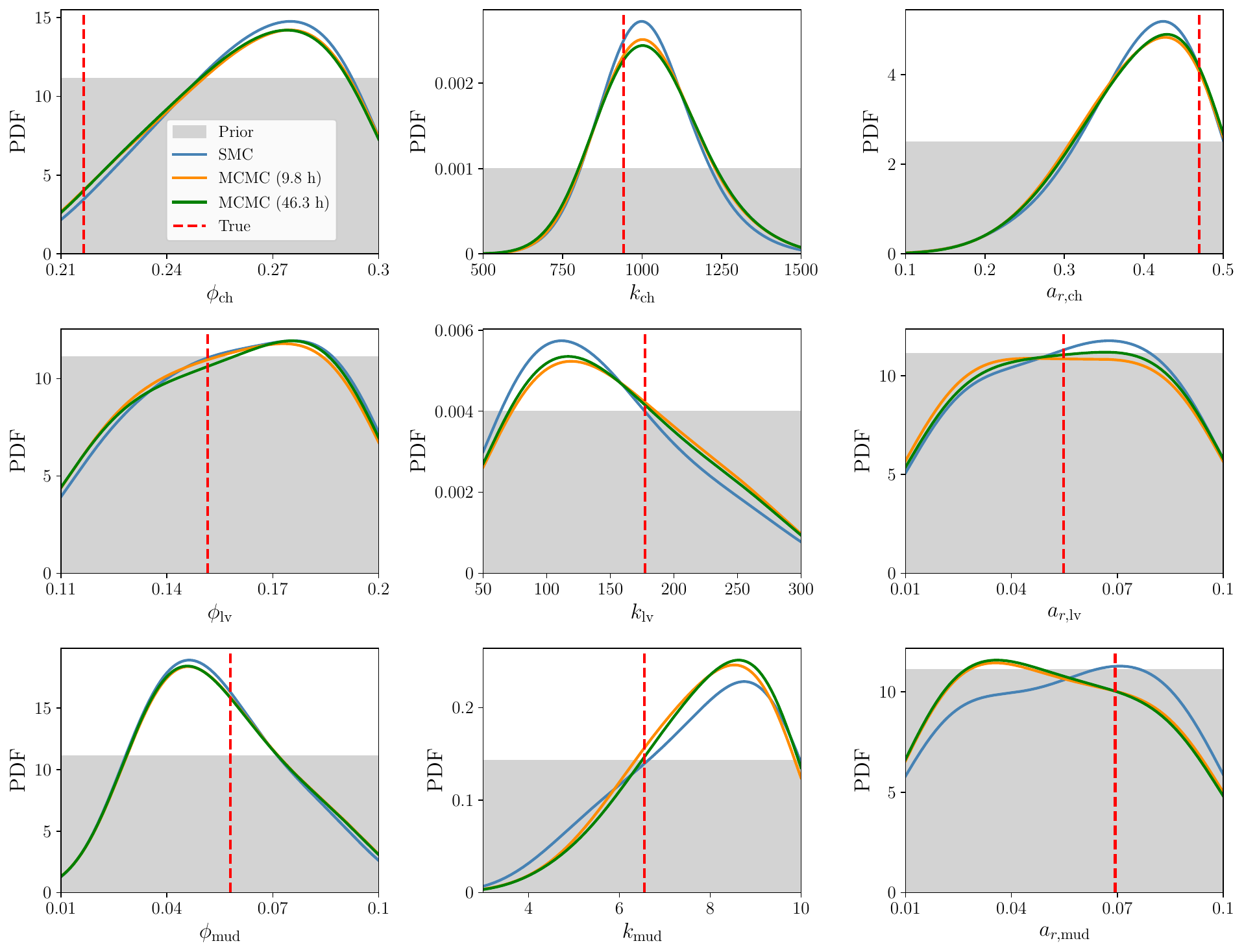}
\caption{Marginal posterior distributions of the nine facies-dependent petrophysical metaparameters in Case~2: porosity $\phi$, permeability $k$, and anisotropy ratio $a_r$ for the channel (top row), levee (middle row), and mud (bottom row) facies. Posterior distributions obtained with SMC are shown in blue. The MCMC distributions at the convergence criteria $\max_j \widehat{R}_j < 1.05$ (9.8~hours) and $\max_j \widehat{R}_j < 1.01$ (46.3~hours) are shown in orange and green, respectively. Gray region denotes the uniform prior, and the red dashed vertical lines indicate the true parameter values.}
\label{fig:tm411-k-phi-marginals}
\end{figure}

The marginal posterior distributions for the nine facies-dependent petrophysical metaparameters, obtained using SMC and MCMC, are compared in Fig.~\ref{fig:tm411-k-phi-marginals}. MCMC results at both 9.8~hours (when $\max_{j}\widehat{R}_j<1.05$ is satisfied) and 46.3~hours (when $\max_{j}\widehat{R}_j<1.01$ is satisfied) are shown. The close agreement between these two sets of results indicates that the marginal distributions are stable under additional MCMC sampling. The SMC distributions are also in close correspondence with the MCMC results. Slight discrepancies are observed in a few cases, most noticeably for $k_{\mathrm{mud}}$ and $a_{r,\mathrm{mud}}$, though overall there is a high level of agreement. Clear uncertainty reduction is achieved for a few metaparameters (e.g., the facies permeabilities), though in other cases very little uncertainty reduction is observed ($\phi_{\mathrm{lv}}$, $a_{r,\mathrm{lv}}$, $a_{r,\mathrm{mud}}$). Interestingly, in the case of $\phi_{\mathrm{ch}}$, the mode is clearly shifted from the true value. This is presumably the correct result, since all three curves are consistent. In any event, the results in Fig.~\ref{fig:tm411-k-phi-marginals} demonstrate that SMC reproduces the MCMC posterior distributions for petrophysical metaparameters showing different degrees of uncertainty reduction.

\begin{figure}[!htbp]
\centering
\includegraphics[width=\linewidth]{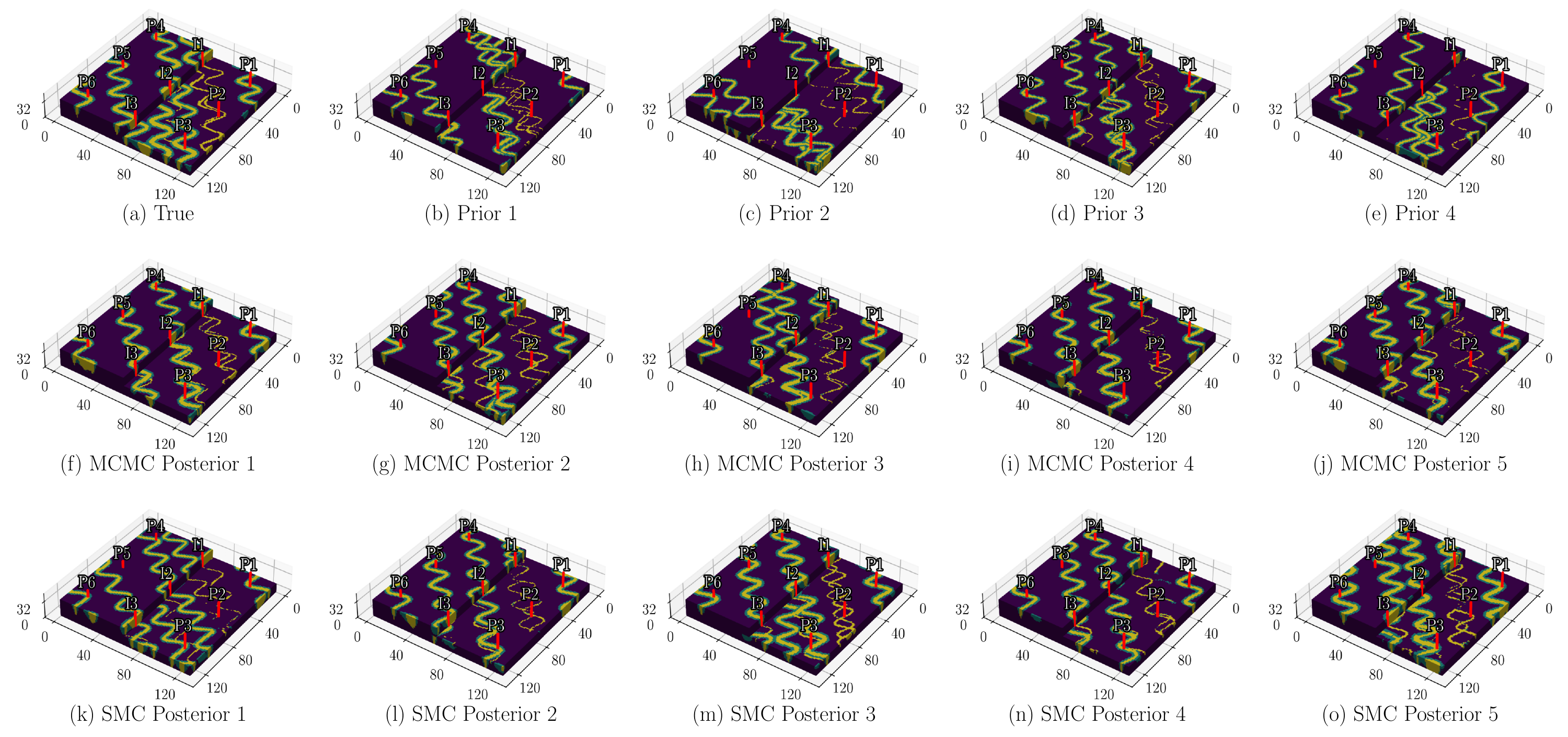}
\caption{Representative 3D facies geomodels for Case~2. Top row shows the true model and four representative prior realizations, and middle and bottom rows show representative MCMC and SMC posterior samples.}
\label{fig:tm411-facies-geomodels}
\end{figure}

Representative 3D facies geomodels for Case~2 are shown in Fig.~\ref{fig:tm411-facies-geomodels}. The true geomodel and four prior realizations are shown in the top row, while the middle and bottom rows show five posterior realizations obtained using MCMC and SMC, respectively. Although all realizations are conditioned to the same hard facies data at the nine wells, the prior geomodels exhibit substantial variability in the spatial distribution of facies. For example, in Prior~2 the channels are oriented about 30$^\circ$ from the $x$-axis, while in Prior~4 they are about 30$^\circ$ from the $y$-axis. The posterior realizations, by contrast, show more consistent channel configurations and closer agreement with the main channel patterns in the true geomodel. Some variability in channel distribution remains among the posterior realizations, particularly away from the wells. This reflects the remaining uncertainty in the spatial facies distribution. The MCMC and SMC posterior geomodels show similar geological features and levels of variability.

\begin{figure}[!htbp]
\centering
\includegraphics[width=\linewidth]{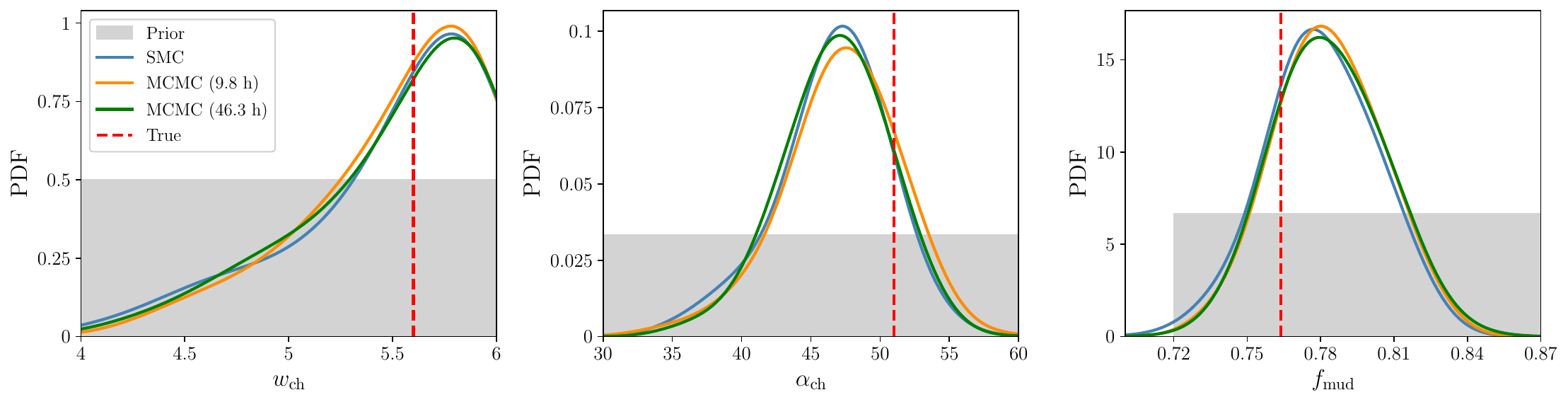}
\caption{Marginal posterior distributions of the three geometric metaparameters for Case~2. Results are shown for SMC (blue) and for MCMC at the satisfaction of $\max_j\widehat{R}_j<1.05$ (9.8~hours, orange) and $\max_j\widehat{R}_j<1.01$ (46.3~hours, green). Uniform priors are in gray and the true parameter values are indicated by the red dashed vertical lines.}
\label{fig:tm411-scenario-marginals}
\end{figure}

The geometric characteristics of the facies realizations are assessed quantitatively through the marginal posterior distributions of $w_{\mathrm{ch}}$, $\alpha_{\mathrm{ch}}$, and $f_{\mathrm{mud}}$, shown in Fig.~\ref{fig:tm411-scenario-marginals}. These geometric metaparameters are not directly updated during data assimilation but are determined from the posterior facies realizations generated by the 3D-LDM. Following the procedure described in \citet{difederico2025ThreeDimensional}, $w_{\mathrm{ch}}$ and $\alpha_{\mathrm{ch}}$ are estimated using a separate convolutional neural network (CNN) trained on Petrel-generated facies realizations and their corresponding object-based modeling (OBM) input parameters. The fraction of mud-facies grid cells ($f_{\mathrm{mud}}$) is computed directly from the realizations. The MCMC posterior distributions at the convergence and stopping points are very similar, and the SMC distributions are in close agreement with the MCMC results. Substantial uncertainty reduction is observed for all three geometric metaparameters. The slight extension of the posterior distribution below the prior lower bound for $f_{\mathrm{mud}}$ arises because this bound applies to the OBM input parameter, whereas $f_{\mathrm{mud}}$ here is computed from the posterior 3D-LDM facies realizations, which are not subject to this constraint. Overall, the results in Fig.~\ref{fig:tm411-scenario-marginals} indicate that the observed well-rate data are informative in terms of channel geometry and facies proportions.

\begin{figure}[!htbp]
\centering
\includegraphics[width=0.95\linewidth]{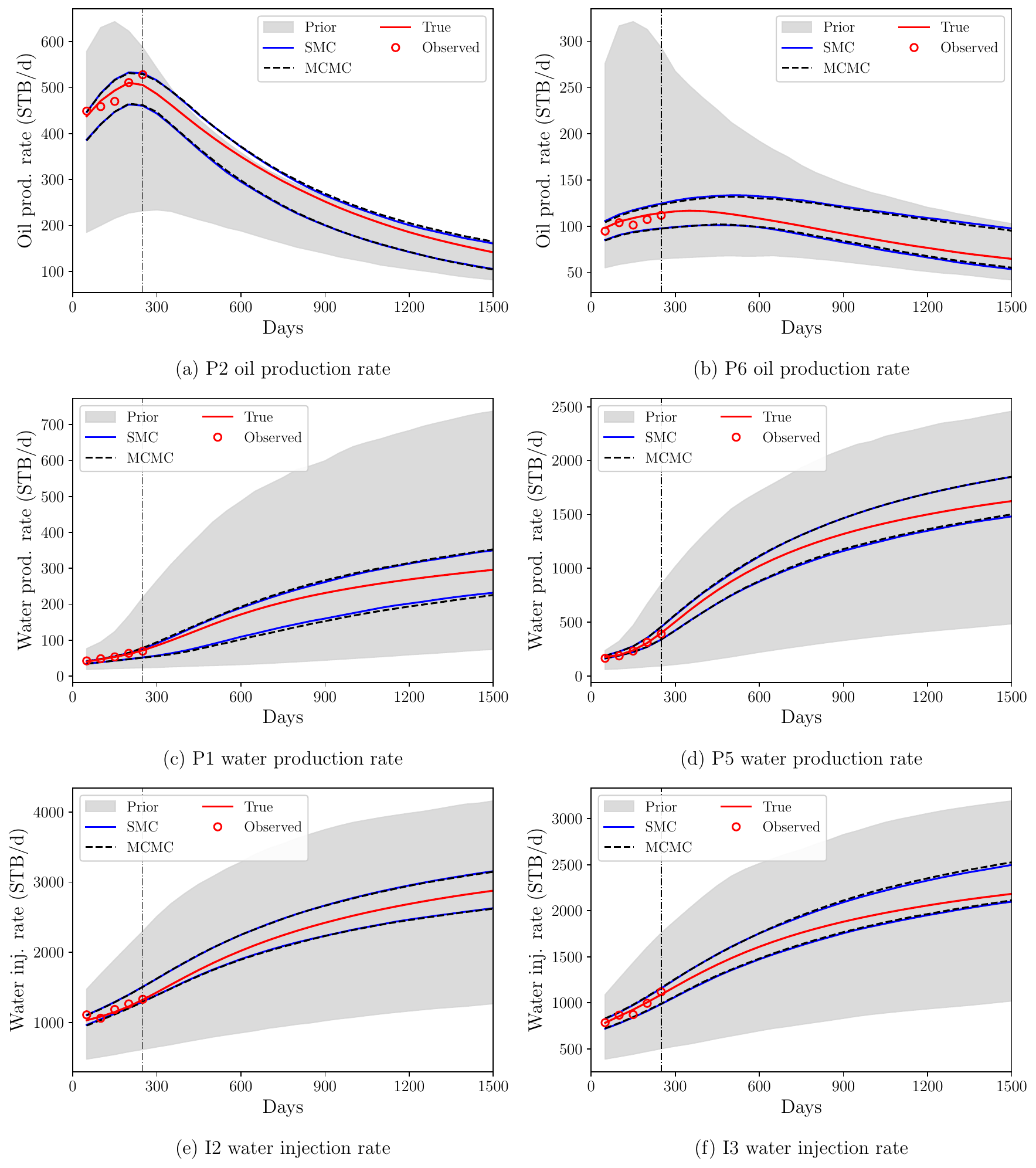}
\caption{Posterior $\mathrm{P}_{10}$--$\mathrm{P}_{90}$ predictions of selected well rates in Case~2. Results are shown for SMC (blue) and MCMC for $\max_{j=1,\ldots,9}\widehat{R}_j<1.05$ (black dashed), along with the prior $\mathrm{P}_{10}$--$\mathrm{P}_{90}$ range (gray), true response (red), and observed data (red circles). Vertical dashed line indicates the end of the data assimilation time window at 250~days.}
\label{fig:tm411-well-rates}
\end{figure}

Posterior predictions for individual well rates are now considered. Fig.~\ref{fig:tm411-well-rates} shows representative well-rate results for oil production, water production, and water injection. Observed data, which include noise, are indicated by the red circles. The prior $\mathrm{P}_{10}$--$\mathrm{P}_{90}$ ranges (shown in gray) are, overall, very broad. Data assimilation using observations through day~250 leads to large uncertainty reduction for all six well responses, and the $\mathrm{P}_{10}$--$\mathrm{P}_{90}$ posterior predictions capture the temporal evolution of the true well-rate responses. The MCMC posterior predictions at 9.8~hours (when $\max_{j=1,\ldots,9}\widehat{R}_j<1.05$ is satisfied) and 46.3~hours (when $\max_{j=1,\ldots,9}\widehat{R}_j<1.01$ is satisfied) are very close, and only the 9.8-hour results are shown. The SMC and MCMC posterior predictions are in excellent agreement over the full 1500-day simulation period. Similar uncertainty reduction and close alignment between SMC and MCMC are observed for field-level oil and water production rates and water injection rates, as shown in Fig.~S9 in SM.

The computational costs of MCMC and SMC for the channelized oil-water system are summarized in Table~\ref{tab:ow-computational-efficiency}. Timings for Case~4 are also included here (detailed results for this case are provided in SM). The timings in Table~\ref{tab:ow-computational-efficiency} indicate MCMC-to-SMC elapsed-time ratios of about 2.5 and 7.0. The MCMC timings and number of function evaluations given here are for the less stringent convergence criterion ($\max_{j=1,\ldots,9}\widehat{R}_j<1.05$). With the more stringent criterion ($\max_{j=1,\ldots,9}\widehat{R}_j<1.01$), the MCMC elapsed time for Case~2 increases to 46.3~hours. For Case~4, the calculation terminates at the 50-hour time limit without satisfying this more stringent criterion. The speedup provided by SMC relative to these longer MCMC elapsed times is about 11.6 in both cases.

\begin{table}[!htbp]
\centering
\small
\caption{Computational cost comparison for MCMC and SMC for the channelized oil-water system (Cases~2 and~4). Elapsed times and numbers of function evaluations are reported for single-GPU calculations.}
\label{tab:ow-computational-efficiency}
\begin{tabular*}{0.95\linewidth}{@{\extracolsep{\fill}}lrrrr@{}}
\toprule
& \multicolumn{2}{c}{Case~2}
& \multicolumn{2}{c}{Case~4} \\
\cmidrule(lr){2-3}\cmidrule(lr){4-5}
Quantity & MCMC & SMC & MCMC & SMC \\
\midrule
Elapsed time (h) & 9.8 & 4.0 & 30.1 & 4.3 \\
Function evaluations
& $638{,}188$ & $361{,}438$
& $1{,}630{,}522$ & $398{,}028$ \\
\bottomrule
\end{tabular*}
\end{table}

As discussed in Section~\ref{sec:3t-sample9}, the population-based structure of SMC makes it well suited for multi-GPU parallelization. We performed additional parallel SMC calculations for Cases~2 and~4 using five NVIDIA A100 GPUs (40~GB each). The parallel SMC runs required 0.84 and 0.94~hours, respectively, corresponding to speedups of approximately 4.8 and 4.6 relative to the single-GPU SMC calculations. Relative to MCMC, these parallel SMC runs provided speedups of about 11.7 and 32.0 for Cases~2 and~4. Comparisons using the sampling efficiency metric described in Section~\ref{sec:3t-sample9} also indicated higher sampling efficiency for SMC in these cases.

Additional results for Case~4 are provided in Figs.~S10--S15 in SM. The synthetic true model for Case~4 has a smaller average channel width and a larger mud fraction than that for Case~2. The SMC and MCMC posterior distributions for the petrophysical and geometric metaparameters for Case~4 are again in close agreement, as are the well and field-rate predictions. Substantial reduction in predictive uncertainty relative to the prior is again observed. 

\FloatBarrier

\section{Concluding remarks}
\label{sec:conclusions}

In this study, we developed an adaptive tempered sequential Monte Carlo (SMC) procedure for data assimilation in high-dimensional subsurface flow systems with hierarchical uncertainty. The method uses noncentered geological parameterizations to represent geomodel realizations in terms of metaparameters and Gaussian latent variables. These two sets of variables, which are independent under the prior, are jointly estimated within a single likelihood-based Bayesian framework. Our SMC method transports a population of particles from the joint prior to the posterior through adaptive likelihood tempering, importance weighting, resampling, and Markov chain Monte Carlo (MCMC) mutation. Temperature increments, which affect how the likelihood term is incorporated in the sampling, are selected using an importance-weight effective sample size criterion. The two sets of variables, which are treated separately (blockwise) in the MCMC mutation step, involve short mutation chains. Deep learning flow surrogates are applied to provide fast forward evaluations. The SMC likelihood term accounts for both measurement error and correlated surrogate model error.

We applied the SMC procedure to two very different 3D subsurface flow systems. The first was a faulted carbon storage system containing three stacked aquifers and two extensive faults. Geological metaparameters characterize the uncertain aquifer and fault properties for this case. Principal component analysis (PCA) was used to represent cell-by-cell heterogeneity in the target storage aquifer. The second case entailed a channelized oil-water system containing sand (channel), levee, and mud facies. A 3D latent diffusion model (3D-LDM) was used to represent variability in channel geometry and facies realizations. The metaparameters in this case characterize the uncertain facies-dependent petrophysical properties. For both systems, SMC was shown to provide posterior metaparameter distributions in close agreement with those obtained using reference MCMC computations. In the carbon storage case, the two methods also produced similar posterior CO$_2$ plume geometries, pressure fields, and saturation and pressure predictions at the monitoring locations. In the oil-water case, the posterior facies realizations and derived geometric metaparameter distributions were comparable, as were the well-by-well and field-rate predictions. Substantial reduction in predictive uncertainty was achieved in both systems. Results for two additional cases -- one for the carbon storage system and one for the oil-water system -- are provided in Supplementary Material. Over the four cases considered, SMC required fewer function evaluations and shorter elapsed times than MCMC, with speedups ranging from about 2.5 to 7.0 for runs that used a single GPU. Parallel SMC calculations using five GPUs (per case) achieved speedups of about 4.6--4.8 relative to single-GPU SMC, and 11.7--32.0 relative to MCMC.

Another important advantage of the proposed SMC procedure is its ability to perform successive (multistage) data assimilation. Specifically, as additional observations become available, SMC is able to update the posterior particle population obtained from the preceding assimilation stage. Conventional MCMC, by contrast, generally requires a new sampling run targeting the posterior conditioned on all available observations. We applied our SMC procedure for two-stage successive data assimilation for the faulted carbon storage system. The total cost for two-stage SMC was 11.5~hours, while the MCMC calculation using all data at once reached its 50-hour time limit without satisfying the convergence criterion. These findings suggest that the SMC method introduced in this study could be very useful in practical cases where new data become available continuously or in batches.

There are several useful directions for future work in this general area. Surrogate model accuracy could be further improved through active learning, with additional high-fidelity training simulations targeted to regions of parameter space associated with larger prediction errors. The successive SMC procedure should be evaluated for cases with multiple data assimilation stages. It should also be applied for sequential decision-making under uncertainty, with updated posterior distributions used to guide subsequent operational decisions. Another topic to be considered is the treatment of model discrepancy, which is often neglected in subsurface history matching. Although surrogate model error was incorporated in the present formulation, discrepancies between the assumed physical model and the actual subsurface system were not considered. Finally, our new SMC procedure should be applied to more realistic systems and other subsurface flow applications, which could include geothermal energy production and freshwater aquifer management.

\section*{CRediT authorship contribution statement}
\textbf{Wenchao Teng}: Conceptualization, Methodology, Software, Formal analysis, Visualization, Writing -- original draft. \textbf{Guido Di Federico}: Conceptualization, Methodology, Formal analysis, Software. \textbf{Louis J.~Durlofsky}: Supervision, Conceptualization, Formal analysis, Resources, Writing -- review \& editing.

\section*{Declaration of generative AI use in the writing process}
In writing the initial draft of this paper, ChatGPT was used to improve wording. The authors reviewed and edited the paper and take full responsibility for its content.

\section*{Declaration of competing interest}
The authors declare that they have no known competing financial interests or personal relationships that could have appeared to influence the work reported in this paper.

\section*{Acknowledgments}
We are grateful to the Stanford Smart Fields Consortium, the Stanford Center for Carbon Storage, and the Stanford Graduate Fellowship in Science and Engineering for financial support. We thank Yifu Han for providing the modified SEAM CO$_2$ geomodel, GEOS simulation files, and the recurrent transformer U-Net code, as well as for useful discussions. We acknowledge the SDSS Center for Computation for providing the HPC resources used in this work.

\section*{Data availability}
The codes developed in this work will be made available upon publication of this paper.

\bibliographystyle{elsarticle-num-names}
\bibliography{references}

\end{document}